\documentclass[
  preprint,
  amsmath,
  amssymb,
  aps,
  prfluids,
  showkeys,
  longbibliography
]{revtex4-2}
\usepackage{graphicx}
\usepackage{etoolbox}
\usepackage{tikz}
\usetikzlibrary{positioning,fit}
\usetikzlibrary{calc,trees,positioning,arrows,chains,shapes.geometric,
  decorations.pathreplacing,decorations.pathmorphing,shapes,
  matrix,shapes.symbols}
\usetikzlibrary{fit}
\usepackage{booktabs}
\usepackage{placeins}
\usepackage[normalem]{ulem}
\begin{document}

\title{A Calibrated Reduced-Order Force Model for Bacterial Hydrodynamics in Free Space and Near a Planar Boundary}

\author{Hoa Nguyen}
\email[Corresponding author: ]{hnguyen5@trinity.edu}
\affiliation{Department of Mathematics, Trinity University, San Antonio, Texas, USA}

\author{William Wallace}
\email{william.c.wallace0@gmail.com}
\affiliation{Department of Mathematics, Trinity University, San Antonio, Texas, USA}

\author{Orrin Shindell}
\email{oshindel@trinity.edu}
\affiliation{Department of Physics and Astronomy, Trinity University, San Antonio, Texas, USA}

\author{Frank Healy}
\email{fhealy@trinity.edu}
\affiliation{Department of Biology, Trinity University, San Antonio, Texas, USA}

\author{Ricardo Cortez}
\email{rcortez@tulane.edu}
\affiliation{Department of Mathematics, Tulane University, New Orleans, Louisiana, USA}

\author{Bruce Rodenborn}
\email{bruce.rodenborn@centre.edu}
\affiliation{Department of Physics, Centre College, Danville, Kentucky, USA}

\begin{abstract}

Accurately resolving the near-field flow generated by many swimming bacteria while
retaining computational efficiency remains challenging. Stokeslet-based models with many force points can capture detailed near-field hydrodynamics but are computationally expensive. Other approaches include far-field and low-order models that use fewer points but sacrifice near-field accuracy. We introduce a reduced-order framework based on the method of regularized Stokeslets that preserves important near-field flow features while using substantially fewer force points. The method replaces the force distribution of a high-fidelity model with forces on a sparse set of points whose strengths are determined by constrained least-squares calibration to the high-fidelity velocity
field of a single bacterium. The calibrated forces yield an approximation that preserves the dominant reference-flow structure while satisfying the force-free and torque-free conditions of self-propelled swimming.
Principal component analysis is then used to represent the phase-dependent variation of the calibrated forces using only a few dominant modes. By fitting the corresponding modal coefficients as continuous functions of flagellar phase, the reduced-order forces can be approximated at any phase of the flagellar cycle. The framework is applied in both free space and near a no-slip planar boundary,
where it preserves the dominant flow structures and captures the wall-induced
redirection of the surrounding fluid. The optimized cell-body regularization parameter depends weakly on wall distance, and the remaining velocity discrepancy is
concentrated primarily near the cell body. By substantially reducing the number of force points, the reduced-order model lowers both the computational cost and memory requirements of velocity-field evaluation, making large-domain simulations of multi-swimmer flow fields more practical.

\end{abstract}


\maketitle

\section{Introduction}

Swimming microorganisms operate in the Stokes flow regime, where inertia is negligible
and long-range fluid interactions govern locomotion, biofilm formation, nutrient transport, and collective behavior~\cite{LaugaPowers2009,Lauga2016,Koch2011}.
Near solid boundaries these interactions become especially important: swimming bacteria can
be hydrodynamically attracted to surfaces, reoriented parallel to boundaries, and driven
into near-wall trajectories that differ substantially from their bulk
motion~\cite{Lauga2006,Berke2008,Spagnolie2012,Drescher2011}. Studying these
phenomena therefore requires computational models that resolve the near-field
flow around individual swimmers while remaining tractable for systems
containing many cells.

Computational models based on the method of regularized Stokeslets~\cite{Cortez2001,MRS} have been used to study helical-flagellum propulsion and bacterial swimming kinematics~\cite{Hyon2012,Rodenborn2013,shindell}. Here, we build on the model of Shindell~\emph{et al.}~\cite{shindell}, which represents an idealized bacterium as a cylindrical cell body propelled by a rotating helical structure. In their work, the size of the regularization parameters or ``blob sizes'' controlling the spatial spread of the regularized forces were calibrated using dynamically similar macroscopic experiments for the cylindrical body and helical flagellum motions. This detailed model provides the reference velocity fields in our work, including both near- and far-field flows throughout the flagellar cycle. This single-helix idealization can model a flagellar bundle, as in \emph{Escherichia coli}, or a single polar flagellum, as in \emph{Pseudomonas aeruginosa}~\cite{LaugaPowers2009,Tian2022} by varying the computational parameters. Here, we use their model to demonstrate construction of a reduced-order model rather than to represent a specific bacterial species. Our work extends naturally to planar no-slip boundaries when combined with the method of images for regularized Stokeslets~\cite{MIRS}.

The surface and flagellum discretization in Shindell~\emph{et al.}~\cite{shindell} comes at a substantial computational cost; the cell body alone
requires more than one thousand force points. The cost of evaluating the velocity field is proportional to the number of bacterial force points and the number of evaluation points. Consequently, increasing either the number of bacteria or the size and resolution of the computational domain substantially increases the computational work and memory requirements, making large-domain, many-bacterium calculations expensive for the detailed model. At the opposite extreme, classical force-dipole and other low-order singularity models~\cite{LaugaPowers2009, lushi2013modeling, hernandez2005transport, hernandez2007fast, hernandez2009dynamics} are computationally inexpensive but do not resolve detailed near-field features, such as the localized rotational flow generated by a rotating helical flagellum, which can be important when modeling closely spaced swimmers. Ishimoto, Gaffney, and Walker~\cite{ishimoto} address this limitation using two regularized Stokeslets and two regularized rotlets, which significantly improves force-dipole descriptions with a small number of singularities.
More recently, Hoover, Boindala, and Cortez~\cite{Hoover2025} developed a
single-particle regularized force-doublet framework for self-propelled
microswimmers, with extensions for rotational flow and planar-wall effects.
Yang et al.~\cite{Yang2026} proposed a compact bacterial-flow representation
combining an anisotropically regularized stresslet with an isotropically
regularized source dipole, and validated the model against experimental and
boundary-element flow fields in free space and near a no-slip wall.

Other approaches reduce regularized-Stokeslet costs through improved
discretization, evaluation, or dimensional reduction. Boundary-element and
nearest-neighbor formulations decouple force degrees of freedom from quadrature
points~\cite{Smith2009,Smith2018}, Richardson extrapolation permits coarser
blob sizes~\cite{Gallagher2021}, and regularized-Stokeslet segments use
continuous piecewise-linear force distributions for slender filaments,
including near a planar no-slip boundary~\cite{Walker2019}. Related
dimensional-reduction approaches have applied Principal Component Analysis (PCA) to microswimmer kinematics
and velocity fields and subsequently represented leading flow modes using
small numbers of regularized Stokeslets~\cite{ISHIMOTO20181}. The choice of
blob size and surface discretization is also important near boundaries.
Nguyen et al.~\cite{Nguyen2025} used theory and macroscopic experiments for
sphere motion near a planar boundary to relate the optimal regularization
scale to the surface discretization and to identify loss of accuracy when the
wall gap becomes insufficiently resolved.

Our strategy is complementary to these approaches. Rather than beginning with
a prescribed low-order singularity representation, modifying the quadrature
rule, or constructing a sparse force distribution through geometric
down-sampling, we prescribe a sparse set of regularized-Stokeslet locations and
determine their force strengths by constrained calibration to the velocity
field generated by a high-fidelity model. 
At each calibration phase, the reduced forces are chosen to approximate the high-fidelity velocity field while satisfying the force-free and torque-free constraints of self-propelled swimming. PCA is then used to represent the phase-dependent variation of the calibrated forces over the flagellar cycle using only a few dominant modes. The corresponding modal coefficients are fitted as continuous functions of phase, allowing the reduced-order forces to be evaluated at any flagellar phase without repeating the calibration.

We extend the free-space calibration framework to bacteria swimming near a planar no-slip boundary over a range of wall distances by replacing the free-space regularized-Stokeslet kernel with the method-of-images kernel for regularized Stokeslets~\cite{MIRS}. The calibrated reduced-order model preserves the dominant phase-dependent force structure while substantially reducing the number of source points and the associated velocity-evaluation cost and storage requirements. This provides an efficient route from high-fidelity regularized-Stokeslet models to tractable flow calculations involving many swimmers.

The methodology is validated against the 1,417-point reference model in free space
and over ten wall distances. Its robustness and broader applicability are further
assessed through continuous-phase reconstruction, out-of-sample phase validation,
and the transformation and assembly of prescribed
multi-swimmer configurations. Together, these tests demonstrate that constrained
inverse calibration can provide compact, phase-dependent regularized-Stokeslet force
representations suitable for large-domain and multi-swimmer flow evaluation.


The remainder of the paper is organized as follows.
Section~\ref{sec:methods} reviews the mathematical foundations of the detailed and
reduced-order models.
Section~\ref{sec:algorithm} describes the calibration procedure and simulation setup.
Section~\ref{sec:results} presents the numerical results, including qualitative validation,
blob-size optimization, spatial error analysis, PCA compression, force-distribution
analysis, and multi-bacterium demonstrations.
Section~\ref{sec:conclusion} summarizes the principal findings and outlines directions for
future work.

\section{Mathematical Methods}\label{sec:methods}

We begin by describing the hydrodynamic model that serves as the high-fidelity
reference throughout this work. The detailed model of Shindell~\emph{et al.}~\cite{shindell}
represents a bacterium as a cylindrical cell body driven by a rotating helical flagellum,
with the body and flagellum discretized by regularized Stokeslet force points.
The reduced-order model is constructed in two stages. First, at each sampled calibration
phase, the detailed force distribution is replaced by a much smaller set of calibrated
regularized-Stokeslet forces chosen to approximate the high-fidelity velocity field while
satisfying the force-free and torque-free constraints of self-propelled swimming. PCA is
then applied to these calibrated forces to obtain a few-mode representation whose smoothly
fitted coefficients permit evaluation at any flagellar phase. 

\subsection{Method of Regularized Stokeslets}\label{sec:MRS}

Both the high-fidelity model and the reduced-order model are
based on the method of regularized Stokeslets (MRS)~\cite{Cortez2001,MRS}. In free-space simulations, the fluid domain is unbounded, and the finite computational grid
used to evaluate the velocity field is centered around the bacterium. The bacterium is oriented with its symmetry axis along the $z$-axis, and the junction between the cell body and flagellum is placed at the origin.

The Reynolds number is negligibly small at the scales relevant to bacterial locomotion, so
 the fluid motion is described by the Stokes equations:
\begin{equation} \label{Eq_MRS}
\mu\,\triangle\mathbf{u}(\mathbf{x}) - \nabla p(\mathbf{x}) = -\mathbf{f}(\mathbf{x}), 
\qquad 
\nabla\cdot\mathbf{u}(\mathbf{x}) = 0.
\end{equation}
Here $\mathbf{u}(\mathbf{x})$ is the fluid velocity, $p(\mathbf{x})$ is the pressure,
$\mu$ is the dynamic viscosity, and $\mathbf{f}(\mathbf{x})$ is the force density
applied to the fluid by the swimmer. The first equation represents the balance between
viscous stress, pressure, and applied force in the absence of inertia, while the second
equation enforces incompressibility.

The MRS removes the singularity of the classical point-force solution by replacing
the Dirac delta distribution with a smooth regularization function
\(\phi_\epsilon\). We use the radially symmetric function
\begin{equation}
\phi_\epsilon(\mathbf{x}-\mathbf{x}_k) =
\frac{15\epsilon^4}{8\pi\,(r_k^2+\epsilon^2)^{7/2}},
\qquad
r_k = \|\mathbf{x}-\mathbf{x}_k\|_2.
\end{equation}
Here \(\mathbf{x}\) is the evaluation point in the fluid, \(\mathbf{x}_k\) is the
location of the \(k\)th regularized-Stokeslet force point, \(r_k\) is the distance
between these points, and \(\epsilon\) is the blob size controlling the spatial spread
of the regularized force. Thus, \(\phi_\epsilon\) replaces the singular point force
with a smooth, radially symmetric force distribution centered at
\(\mathbf{x}_k\). This regularization produces a bounded velocity kernel, so the
velocity remains finite at the force locations along discretized curves or
surfaces~\cite{Cortez2001,MRS}.

For $N_y$ force points $\{\mathbf{x}_k\}$ and $N_x$ evaluation points, the linearity of the
Stokes equations gives the discrete relation
\begin{equation}
\mathbf{U} = M\mathbf{F},
\end{equation}
where $\mathbf{U}\in\mathbb{R}^{3N_x}$ and $\mathbf{F}\in\mathbb{R}^{3N_y}$ are the
stacked velocity and force vectors, and
$M\in\mathbb{R}^{3N_x\times 3N_y}$ is the regularized-Stokeslet matrix. In the detailed model, the force points coincide with the collocation points, and the unknown forces and rigid-body velocities are obtained by solving the augmented square system described in 
Section~\ref{sec:constraints}. The resulting discrete forces are then used in the MRS to evaluate the fluid velocity on the surrounding computational grid. In the reduced-order model, the same flow field is approximated using a much smaller set of force points, whose strengths are determined by constrained inverse calibration to the detailed-model velocity field on this grid, as described in Section~\ref{sec:calibration}.

\subsection{Method of Images for Regularized Stokeslets}\label{sec:MIRS}

The no-slip condition on a planar boundary is enforced using the method of images for
regularized Stokeslets (MIRS)~\cite{MIRS}. The MIRS uses the same coordinate convention as in the
free-space simulations with the planar wall located at $x=h<0$, so the fluid domain occupies
the region $x>h$. The computational grid extends from the wall into the fluid region in
the positive $x$-direction.

The MIRS places a reflected point across the wall for each regularized Stokeslet located in the fluid domain. This image system consists of an
image Stokeslet, a Stokeslet doublet, a potential dipole, and two rotlets, which together
cancel the velocity at the wall for each regularized Stokeslet~\cite{MIRS}. 
The linearity of the Stokes equations gives a discrete
matrix relation between the force vector and the velocity vector,
\begin{equation}
\mathbf{U} = M_{\mathrm{MIRS}}\mathbf{F},
\end{equation}
where $\mathbf{U}\in\mathbb{R}^{3N_x}$ and $\mathbf{F}\in\mathbb{R}^{3N_y}$ are the
stacked velocity and force vectors, and
$M_{\mathrm{MIRS}}\in\mathbb{R}^{3N_x\times 3N_y}$ is the method-of-images
regularized-Stokeslet matrix. The only change from the free-space MRS system is that each
matrix entry is assembled using the wall-corrected image-system kernel instead of the
free-space regularized Stokeslet kernel.

We require the reduced-order cell-body blob size, $\epsilon_{r,c}$, to remain
smaller than the minimum distance from the cell body to the wall~\cite{Zheng}
for nine of the ten near-wall configurations. When $\epsilon_{r,c}$ is
comparable to or larger than the wall--body gap, a substantial portion of the
regularized forcing extends across the wall, weakening its interpretation as
a force distribution acting within the fluid surrounding the body. The
closest-wall configuration is the sole exception to this blob-size--gap
condition and is used as a numerical stress test, as discussed in
Section~\ref{sec:decay}.

\subsection{Force-Free and Torque-Free Constraints}\label{sec:constraints}

A swimming bacterium at low Reynolds number is force-free and torque-free. In the
detailed model of Shindell~\emph{et al.}~\cite{shindell}, these constraints are imposed
by augmenting the regularized-Stokeslet system with six additional equations: three for
zero net force and three for zero net torque. The augmented system simultaneously
determines the translational velocity $\mathbf{u}_b$, the angular velocity
$\boldsymbol{\Omega}_b$ of the cell body, and the internal force vectors
$\mathbf{f}_k$:
\begin{equation}\label{Eq_MRS_force_torque_free}
\begin{split}
\tilde{\mathbf{u}}(\mathbf{x}_j) &=
\frac{1}{8\pi\mu}\sum_{k=1}^{N_y}
    G_\epsilon(\mathbf{x}_j,\mathbf{x}_k)\mathbf{f}_k
    - \mathbf{u}_b
    - \boldsymbol{\Omega}_b\times(\mathbf{x}_j-\mathbf{x}_r),
    \quad j=1,\ldots,N_x, \\[4pt]
\sum_{k=1}^{N_y}\mathbf{f}_k &= \mathbf{0}, \qquad
\sum_{k=1}^{N_y}(\mathbf{x}_k-\mathbf{x}_r)\times\mathbf{f}_k = \mathbf{0}.
\end{split}
\end{equation}
Here $G_\epsilon$ denotes the free-space regularized-Stokeslet kernel in MRS or the
wall-corrected image-system kernel in MIRS. The point $\mathbf{x}_r$ is the junction
between the cell body and flagellum and serves as the motor location and torque reference
point.

The prescribed velocity $\tilde{\mathbf{u}}(\mathbf{x}_j)$ represents the velocity at the discretization points relative to $\mathbf{x}_r$ measured in the body frame, which is zero on the cell body before the unknown rigid-body translation and rotation
of the bacterium are added. The body frame velocity for points on the flagellum is determined by the motor rotation:
\begin{equation}
\tilde{\mathbf{u}}(\mathbf{x}_j)
=
\boldsymbol{\Omega}_m \times (\mathbf{x}_j-\mathbf{x}_r),
\qquad \mathbf{x}_j \ \text{on the flagellum},
\end{equation}
where $\boldsymbol{\Omega}_m$ is the motor angular velocity vector. Thus, the flagellum
rotates relative to the cell body, while the unknown quantities $\mathbf{u}_b$ and
$\boldsymbol{\Omega}_b$ determine the resulting rigid-body swimming motion required
to satisfy the force-free and torque-free constraints.

\subsection{Constrained Inverse Calibration of Reduced Forces}
\label{sec:calibration}

The reduced-order model replaces the large set of force points in the 
high-fidelity model~\cite{shindell} with \(m\) phase-dependent regularized-Stokeslet locations. The force strengths at these locations are not prescribed;
they are determined by constrained inverse calibration. 
At a given flagellar phase, let
$\mathbf{U}_D\in\mathbb{R}^{3N_x}$ denote the velocity field generated by the detailed
model at $N_x$ points on the computational grid surrounding the bacterium.
Let
$
M_R\in\mathbb{R}^{3N_x\times 3m}
$
denote the reduced regularized-Stokeslet matrix that maps the stacked reduced force vector
\(\mathbf{F}_R\in\mathbb{R}^{3m}\) to the corresponding reduced-order velocity field
\(M_R\mathbf{F}_R\) at the \(N_x\) grid points. In free space, the entries of $M_R$ are
computed using the MRS kernel. Near a planar no-slip boundary, the entries of $M_R$
are computed using the MIRS image-system kernel. In both cases, \(M_R\) is determined by the
phase-specific reduced-order point locations, the evaluation grid, and the blob sizes assigned to the
cell-body and flagellar points. 

At each sampled flagellar phase, the reduced force vector $\mathbf{F}_R$ is determined by the
constrained least-squares problem
\begin{equation}
\label{eq:constrained_ls}
\min_{\mathbf{F}_R\in\mathbb{R}^{3m}}
\left\|M_R\mathbf{F}_R-\mathbf{U}_D\right\|_2^2
\qquad
\text{subject to}
\qquad
\mathcal{C}\mathbf{F}_R=\mathbf{0},
\end{equation}
where the constraint matrix $\mathcal{C}\in\mathbb{R}^{6\times 3m}$ enforces zero net force and
zero net torque:
\begin{equation}
\label{eq:calibration_constraints}
\sum_{\ell=1}^{m}\mathbf{f}_\ell=\mathbf{0},
\qquad
\sum_{\ell=1}^{m}(\mathbf{y}_\ell-\mathbf{x}_r)\times\mathbf{f}_\ell=\mathbf{0}.
\end{equation}
Here $\mathbf{y}_\ell$ is the position of the $\ell$th reduced force point,
$\mathbf{f}_\ell$ is its associated force vector, and $\mathbf{x}_r$ is the reference point,
taken to be the junction between the cell body and flagellum.

Thus, the same algebraic calibration problem is used in free space and near a wall; the
entries of \(M_R\) change through the choice of hydrodynamic kernel and, in the near-wall
case, through the wall position. This formulation allows the reduced force distribution to be chosen 
so that it provides the least-squares approximation of the high-fidelity
velocity field while satisfying the force-free and torque-free constraints of
self-propelled swimming.

The equality constraint, $\mathcal{C}\mathbf{F}_R=\mathbf{0}$ in Eq.~\eqref{eq:constrained_ls}, was eliminated using an orthonormal null-space
basis, \(Z\), satisfying \(\mathcal{C}Z=\mathbf 0\). Writing
\(\mathbf F_R=Z\mathbf a\) therefore gives
\(\mathcal{C}\mathbf F_R=\mathcal{C}Z\mathbf a=\mathbf 0\) for every
\(\mathbf a\), enforcing the equality constraints identically and converting
Eq.~\eqref{eq:constrained_ls} to the unconstrained least-squares problem
\begin{equation}\label{eq:null_space_calibration}
\min_{\mathbf a}
\left\|M_RZ\mathbf a-\mathbf U_D\right\|_2^2.
\end{equation}
To avoid storing the full calibration matrix, the rows associated with the
evaluation grid were processed in blocks. For block \(b\), let \(M_{R,b}\)
and \(\mathbf U_{D,b}\) denote the corresponding matrix rows and reference
velocities. The reduced normal equation quantities were accumulated as
\begin{equation}\label{eq:block_normal_equations}
G=\sum_b (M_{R,b}Z)^T(M_{R,b}Z),
\qquad
\mathbf g=\sum_b (M_{R,b}Z)^T\mathbf U_{D,b},
\end{equation}
after which \(G\mathbf a=\mathbf g\) was solved and
\(\mathbf F_R=Z\mathbf a\) was reconstructed. No inequality constraints,
force bounds, weighting, or additional regularization terms were used, so
each Cartesian velocity component at each grid point contributed equally to
the objective. The blockwise normal equation formulation avoids storing the
full calibration matrix, whereas direct use of MATLAB's \texttt{lsqlin} would
require assembling and storing that matrix. 
Its implementation was also verified directly against
the original constrained least-squares implementation on a smaller computational
grid, as described in Section~\ref{sec:algorithm}.
The force-free and torque-free residuals were evaluated
after calibration to verify they satisfied the constraints to numerical precision.

\subsection{Principal Component Analysis}\label{sec:PCA}

The model was calibrated at $n=16$ evenly spaced flagellar phases within one flagellar cycle (Section~\ref{comp_procedure}), and
Principal Component Analysis (PCA) is applied to the resulting force ensemble to identify
a compact few-mode representation, in which only a few principal components are retained
to approximate the phase-dependent variation of the calibrated forces.

For each Cartesian force component \(i=1,2,3\), let
\(\psi_i\in\mathbb{R}^{n\times m}\) collect the \(i\)th force component over all
\(n\) sampled phases and \(m\) reduced-order force points, with one row corresponding
to each phase. Let \(\bar{\psi}_i\in\mathbb{R}^{1\times m}\) denote the phase-mean
row and \(\mathbf{1}_n\in\mathbb{R}^{n}\) the column vector of ones. Defining the
centered force matrix
\[
X_i=\psi_i-\mathbf{1}_n\bar{\psi}_i,
\]
we compute the unnormalized covariance matrix separately for each Cartesian force
component as
\begin{equation}\label{eq:PCA_cov}
C_i=X_i^T X_i
=
(\psi_i-\mathbf{1}_n\bar{\psi}_i)^T
(\psi_i-\mathbf{1}_n\bar{\psi}_i)
\in\mathbb{R}^{m\times m}.
\end{equation}

Because the columns of \(X_i\) have zero mean,
\(\mathbf{1}_n^T X_i=\mathbf{0}\), or equivalently, the sum of the rows of
\(X_i\) is the zero vector. Thus, the \(n\) rows are linearly dependent and
\(\operatorname{rank}(X_i)\le n-1\). Since
\(\operatorname{rank}(X_i^T X_i)=\operatorname{rank}(X_i)\), it follows that
\(\operatorname{rank}(C_i)\le n-1\). For the \(n=16\) sampled phases used here,
at most 15 PCA modes can therefore have nonzero variance, even though each
force-component vector contains \(m=46\) entries.

The eigenvectors \(\mathbf{v}_{i,k}\) of \(C_i\), where \(k\) indexes the PCA
mode for force component \(i\), define spatial force patterns across the \(m\)
reduced-order force points. The corresponding eigenvalues
\(\lambda_{i,k}\) quantify the phase-dependent force variation associated with
these patterns. We omit the conventional normalization factor \(1/(n-1)\)
because it rescales all eigenvalues by the same constant and therefore does not
affect the PCA modes, their ordering, or the fraction of variance explained by
each mode,
\[
\frac{\lambda_{i,k}}
{\sum_{\ell=1}^{m}\lambda_{i,\ell}}.
\]


The force component at phase \(p\) is approximated by the truncated PCA expansion
\begin{equation}\label{eq:PCA}
\psi_{i,p} \approx \bar{\psi}_i + \sum_{k=1}^{q_i} B_{i,k}(p)\,\mathbf{v}_{i,k},
\end{equation}
where the phase-dependent modal coefficients are obtained by projection,
\[
B_{i,k}(p)
=
\left(\psi_{i,p}-\bar{\psi}_i\right)\cdot \mathbf{v}_{i,k},
\]
and \(q_i\) is the number of retained modes for force component \(i\).
Here \(F_1\), \(F_2\), and \(F_3\) denote the Cartesian force components in the
\(x\)-, \(y\)-, and \(z\)-directions, respectively. Since the bacterium is oriented along
the \(z\)-axis, \(F_1\) and \(F_2\) are lateral force components, while \(F_3\) is the axial
force component. We use \(q_1=q_2=2\) for the lateral components \(F_1\) and \(F_2\),
and \(q_3=1\) for the axial component \(F_3\). The justification for this mode-count
choice is presented in Section~\ref{sec:pca} where the spectra and constraint
residuals are examined across all eleven configurations, and velocity convergence with
the number of axial modes is tested at three representative wall distances.

PCA has previously been used to obtain low-dimensional representations of flagellar kinematics and microswimmer velocity fields, with leading flow modes further approximated using regularized point forces~\cite{ShapeMode,ISHIMOTO20181}. Here, in contrast, PCA is applied to the phase-dependent calibrated force distributions themselves, after the sparse forces have been determined by the constrained inverse problem in Eq.~\eqref{eq:constrained_ls}. The discrete coefficients \(B_{i,k}(p)\) are then fit smoothly over the flagellar
cycle. Writing \(\theta=2\pi p\) for phase fraction \(p\in[0,1)\), the lateral coefficients are fit by
least squares as
\begin{equation}
 B_{i,k}(\theta)
 =a_{i,k}\sin\theta+b_{i,k}\cos\theta+c_{i,k},
 \qquad i=1,2.
\end{equation}
A periodic cubic interpolant is used for \(F_3\). The periodic boundary conditions enforce matching
coefficient value, slope, and curvature at the cycle seam. These fits provide continuous-phase force approximations that can be evaluated
at any flagellar phase without repeating the calibration. The modal truncation and continuous coefficient fitting may violate the
force-free and torque-free constraints, so the corresponding residuals of the
reconstructed forces and torques are evaluated in Section~\ref{sec:pca}.

\subsection{Error Quantification}\label{sec:RMSE}

The accuracy of the reduced-order model is assessed using the root-mean-square error
(RMSE) between the fluid speeds generated by the detailed and reduced-order models.
The RMSE is defined as
\begin{equation}\label{RMSE_equation}
\mathrm{RMSE} = \sqrt{\frac{1}{N}\sum_{j=1}^N
    \bigl(I_{D,j} - I_{R,j}\bigr)^2},
\end{equation}
where $N$ is the number of grid points,  $j$ indexes the grid point, \(I_{D,j}\) is the detailed model fluid speed, and \(I_{R,j}\) is the reduced-order model's fluid speed at the $jth$ grid point. Grid points inside the cell body are
excluded from the RMSE calculation. The RMSE has units of \(\mu\mathrm{m}/\mathrm{s}\) and provides a
 measure of the discrepancy in speed between the two models, which we use to optimize the reduced-order model in two ways: 
 
1) The \emph{three-dimensional RMSE} is computed using every grid point outside the cell body; the complete grid contains
$201\times31\times99$ points and provides a measure of the overall agreement
between the detailed and reduced-order models.

2) The \emph{$x$-slice RMSE} is computed
on a single plane at fixed $x$, providing an error measure as a function of distance from the cell body and the wall.  

The cell body blob size, \(\epsilon_{r,c}\), controls the spatial spread of the regularized cell body forces and
therefore affects both the reduced-order matrix \(M_R\) and the calibrated force vector
\(\mathbf{F}_R\). We therefore treat \(\epsilon_{r,c}\) as a scalar calibration parameter.
For each candidate value of \(\epsilon_{r,c}\), the corresponding reduced-order matrix is
assembled, the reduced forces are recomputed by solving the constrained least-squares
problem in Eq.~\eqref{eq:constrained_ls}, and the resulting velocity field is compared
with the detailed velocity field using the \emph{three-dimensional RMSE} defined in
Eq.~\eqref{RMSE_equation}. The optimal blob size is then identified in Section~\ref{sec:blob} as the value of
\(\epsilon_{r,c}^*\) that minimizes the \emph{three-dimensional RMSE}. After this selection, the \emph{\(x\)-slice RMSE} is used in
Section~\ref{sec:decay} to analyze the spatial distribution of the error
and to distinguish near-field and far-field contributions.

The reduced-order model is also evaluated
at flagellar phases excluded from calibration as a complementary robustness check. The metric definitions,
validation procedure, and comparison between the calibration and
out-of-sample results are provided in
the Appendix: Out-of-Sample Phase Validation. This out-of-sample phase
validation reports three quantities: (i) the RMSE between the magnitudes of the
cycle-averaged velocity fields, hereafter called the cycle-averaged speed RMSE,
obtained by applying Eq.~\eqref{RMSE_equation} after averaging the velocity components
over the 16 test phases; (ii) the phasewise speed RMSE, obtained by
applying the same equation separately at each test phase; and (iii) a
dimensionless relative vector error that measures discrepancies in both the
magnitude and direction of the velocity field. The two speed-RMSE quantities
are consistent with the primary error definition used in the main analysis,
whereas the relative vector error provides an additional direction-sensitive
diagnostic. The out-of-sample results are used only to evaluate the calibrated
model; no test-phase data or error values are used to refit the force
coefficients or alter the selected blob sizes. 

\section{Algorithm and Model Setup}\label{sec:algorithm}

\subsection{Computational Procedure}\label{comp_procedure}

The preceding sections describe the individual components of the reduced-order framework. For clarity, we summarize here how these components are assembled into the computational workflow used in the simulations. The framework consists of two stages: an \emph{offline calibration} stage and an \emph{online evaluation} stage (Figure~\ref{fig:process}). In the reduced-order workflow, the detailed model is used during the
\emph{offline calibration} stage to generate reference velocity fields at 16 flagellar
phases, and the reduced-order forces are calibrated to match those fields while satisfying
the force-free and torque-free constraints. Additional detailed-model solutions at
out-of-sample phases are used only for validation and are not part of the calibration
or online evaluation procedure.
The calibrated forces over the sampled phases are compressed by principal component analysis, and the fitted PCA representation can be evaluated at any desired flagellar phase. The phase-specific reduced geometry is generated and the sparse force vector is reconstructed from the fitted PCA coefficients for a given phase. The velocity field is then obtained by applying the corresponding MRS or MIRS evaluation operator to this force vector. This  operation generally requires a memory intensive matrix–vector product, but in our work the targets are processed in blocks to limit memory. 

The detailed model of
Shindell~\emph{et al.}~\cite{shindell} is solved using the augmented regularized-Stokeslet
system described in Section~\ref{sec:constraints} for each flagellar phase. This solve determines the detailed
force distribution together with the rigid-body swimming and angular velocities of the cell
body.  The resulting detailed forces are then used to evaluate the reference velocity field
$\mathbf U_D$ on the surrounding computational grid. In free space this evaluation uses
the MRS kernel, while near a planar wall it uses the MIRS kernel described in
Sections~\ref{sec:MRS} and~\ref{sec:MIRS}.

The reduced-order model uses the same target velocity field $\mathbf U_D$, but replaces
the detailed force distribution by forces on a much smaller set of prescribed points.  For
each sampled phase, these reduced forces are determined by the constrained inverse
calibration problem in Section~\ref{sec:calibration}.  This step chooses the sparse force distribution that best approximates the
detailed velocity field on the grid while enforcing zero net force and zero
net torque. 

To reduce the memory requirements of these calculations, OpenAI ChatGPT (GPT-5.6 Sol) was used to assist with modifications to the MATLAB implementation. The detailed-model velocity field was evaluated blockwise, and the reduced-order calibration replaced \texttt{lsqlin} with null-space
elimination of the force- and torque-free constraints followed by blockwise accumulation of the normal equations. These changes affect only the numerical implementation, not the underlying model or constrained least-squares problem. The revised code was independently verified against the original implementation
on a smaller $41\times21\times50$ grid: for all 16 phases in free space and near a representative wall, detailed-model differences were at floating-point roundoff, while reduced-order force and cycle-averaged velocity differences were approximately $10^{-9}$ and $10^{-10}$ in relative terms, respectively.


The calibration grid is the complete Cartesian
grid, including points inside the swimmer geometry. Values at these points are
the smooth mathematical continuation of the regularized-Stokeslet field, not
physical fluid measurements. We retain these continuation values as additional
unweighted equations in the overdetermined inverse problem. This choice also avoids a
geometry-dependent interior-classification step, which becomes ambiguous for a
slender flagellum and increasingly difficult for more complicated swimmer
geometries. Reported velocity errors, in contrast, exclude points inside the
cell body, as described in Section~\ref{sec:RMSE}.

Repeating the calibration over the $n=16$ phases
produces the phase-dependent reduced-force ensemble used in the PCA construction of
Section~\ref{sec:PCA}.
Once the PCA force representation has been constructed, the reduced-order
model can be evaluated at any desired flagellar phase. If
\(\mathbf F_{\mathrm{PCA}}(\theta(t))\) denotes the force vector assembled from the
PCA approximation at angular phase \(\theta(t)\), then the velocity field is
obtained by applying the appropriate MRS or MIRS evaluation matrix to
\(\mathbf F_{\mathrm{PCA}}(\theta(t))\).  
For a fixed target grid and unchanged source geometry, a previously assembled
evaluation matrix can be reused. The flagellar source locations vary
with phase, so evaluating a new phase generally requires assembling the corresponding
phase-specific matrix unless that matrix has been precomputed and cached. A new
matrix is also required when the target grid changes.

\tikzset{
  block/.style={
    rectangle,
    draw,
    text width=16em,
    text centered,
    rounded corners,
    minimum height=4em
  },
  arrow/.style={
    ->,
    thick
  }
}

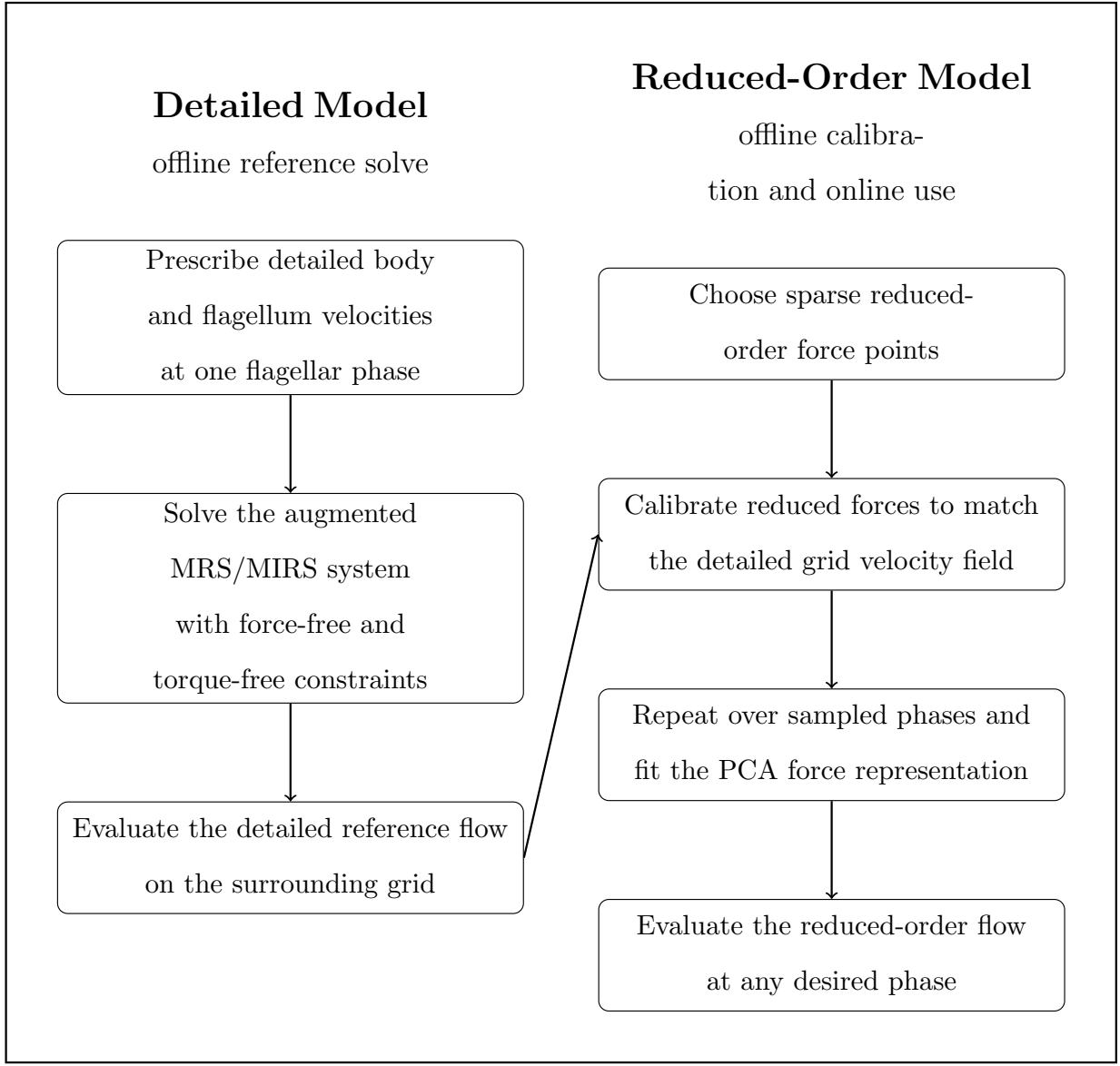
\begin{figure}[!htbp]
    \centering
    \resizebox{\textwidth}{!}{
        \begin{tikzpicture}

            \node (title_d) [text centered, text width=14em]
                {\large{\bfseries{Detailed Model}} \\ \normalsize{offline reference solve}};

            \node (n1) [block, below=0.8cm of title_d]
                {Prescribe detailed body and flagellum velocities\\at one flagellar phase};

            \node (n2) [block, below=1.35cm of n1]
                {Solve the augmented MRS/MIRS system\\with force-free and torque-free constraints};

            \node (n3) [block, below=1.35cm of n2]
                {Evaluate the detailed reference flow\\on the surrounding grid};

            \node (title_r) [text centered, text width=15em, right=1.6cm of title_d]
                {\large{\bfseries{Reduced-Order Model}} \\ \normalsize{offline calibration and online use}};

            \node (n4) [block, below=0.8cm of title_r]
                {Choose sparse reduced-order force points};

            \node (n5) [block, below=1.35cm of n4]
                {Calibrate reduced forces to match\\the detailed grid velocity field};

            \node (n6) [block, below=1.35cm of n5]
                {Repeat over sampled phases and\\fit the PCA force representation};

            \node (n7) [block, below=1.35cm of n6]
                {Evaluate the reduced-order flow\\at any desired phase};

            \draw [arrow] (n1) -- (n2);
            \draw [arrow] (n2) -- (n3);

            \draw [arrow] (n4) -- (n5);
            \draw [arrow] (n5) -- (n6);
            \draw [arrow] (n6) -- (n7);

            \draw [arrow] (n3.east) -- (n5.west);

            \node[
                draw=black,
                thick,
                fit=(title_d)(title_r)(n1)(n2)(n3)(n4)(n5)(n6)(n7),
                inner sep=20pt
            ] {};

        \end{tikzpicture}
    }
\caption{Calibrated reduced-order workflow. During the offline stage, detailed-model
velocity fields at sampled phases are used to calibrate sparse forces subject to the
force-free and torque-free constraints; PCA then compresses their phase dependence. During
the online stage, fitted PCA coefficients reconstruct the sparse force distribution at any
phase, allowing velocity evaluation without repeating the detailed-model solve.
Sections~\ref{sec:MRS}--\ref{sec:PCA} give the mathematical formulation.}
\label{fig:process}
\end{figure}
\FloatBarrier

By the linearity of Stokes flow, for fixed swimmer geometry, wall distance, blob sizes,
and reduced force locations, the calibrated forces scale linearly with the motor rotation
frequency; thus, a calibration performed at \(f_0=154\,\mathrm{Hz}\) can be reused at
another frequency \(f\) by multiplying the calibrated forces by \(f/f_0\).
The same scaling applies to the PCA mean forces and modal coefficients.

Let \(t\) denote physical time, with \(t=0\) chosen as the phase origin, and let
\(f_r\) denote the flagellar rotation frequency. The phase fraction within the current
flagellar cycle is
$
p(t)=f_r t-\lfloor f_r t\rfloor,
$
with \(p(t)\in[0,1)\), and the corresponding angular phase is
$
\theta(t)=2\pi p(t).
$
The PCA coefficient fits, constructed from \(n=16\) equally spaced calibration phases,
can then be evaluated at the phase \(\theta(t)\) corresponding to any simulation time.


\subsection{Model Parameters}\label{sec:params}

Tables~\ref{table:detailed_parameters} and~\ref{table:reduced_parameters}
list the parameters for the detailed and reduced-order models, respectively.
Unless otherwise stated, lengths are reported in \(\mu\mathrm{m}\) and
velocities in \(\mu\mathrm{m}/\mathrm{s}\). Using the corresponding
consistent unit system for the dynamic viscosity, forces are reported in
\(\mathrm{fN}\) (equivalently \(\times10^{-3}\,\mathrm{pN}\)) and torques in
\(\mathrm{fN}\,\mu\mathrm{m}\) (equivalently
\(\times10^{-3}\,\mathrm{pN}\,\mu\mathrm{m}\)).

\begin{table}[!htbp]
    \caption{Parameter values for the detailed model \cite{shindell}.}
    \setlength{\tabcolsep}{10pt}
    \label{table:detailed_parameters}
    \begin{tabular}{lccc}
        \toprule
        \textbf{Parameter} & \textbf{Value} & \textbf{Unit} & \textbf{Description} \\
        \midrule
        $\mu$ & $0.93\times10^{-3}$ & $\mathrm{Pa{\cdot}s}$ & Dynamic viscosity \\
        \midrule
        \multicolumn{4}{l}{Cell body} \\
        \qquad$\ell$ & 2.5 & $\mu\mathrm{m}$ & Length \\
        \qquad$r$ & 0.44 & $\mu\mathrm{m}$ & Radius \\
        \qquad$\gamma_c$ & $6.4$ & & Optimal discretization factor \\
        \qquad$ds_c$ & $0.096$ & $\mu\mathrm{m}$ & Discretization size \\
        \qquad$\epsilon_{d,c} = ds_c/\gamma_c$ & $0.015$ & $\mu\mathrm{m}$ & Blob size \\
        \qquad$N_{d,c}$ & $1046$ & & Points on cell body \\
        \midrule
        \multicolumn{4}{l}{Flagellum} \\
        \qquad$L$ & $8.3$ & $\mu\mathrm{m}$ & Axial length \\
        \qquad$\lambda$ & $2.22$ & $\mu\mathrm{m}$ & Wavelength \\
        \qquad$R$ & $0.2$ & $\mu\mathrm{m}$ & Radius \\
        \qquad$a$ & $0.012$ & $\mu\mathrm{m}$ & Filament radius \\
        \qquad$f_r=\Omega_m/(2\pi)$ & $154$ & \(\mathrm{Hz}\) & Motor frequency \\
        \qquad$\gamma_f$ & $2.139$ & & Optimal discretization factor \\
        \qquad$ds_f = \epsilon_{d,f}$ & $0.026$ & $\mu\mathrm{m}$ & Blob size \\
        \qquad$N_{d,f}$ & $371$ & & Points on flagellum \\
        \bottomrule
    \end{tabular}
\end{table}

\begin{table}[!htbp]
    \caption{Parameter values for the reduced-order model.}
    \setlength{\tabcolsep}{8pt}
    \label{table:reduced_parameters}
    \begin{tabular}{lccc}
        \toprule
        \textbf{Parameter} & \textbf{Value} & \textbf{Unit} & \textbf{Description} \\
        \midrule
        \multicolumn{4}{l}{Cell body} \\
        \qquad$\epsilon_{r,c}^{0}$ & $0.02125$ & $\mu\mathrm{m}$ & Initial cell-body blob size \\
        \qquad$\epsilon_{r,c}^{*}$ & $0.18$ & $\mu\mathrm{m}$ & Optimized cell-body blob size (Section~\ref{sec:blob}) \\
        \qquad$N_{r,c}$ & $25$ & & Points on cell body \\
        \midrule
        \multicolumn{4}{l}{Flagellum} \\
        \qquad$\epsilon_f$ & $0.0385$ & $\mu\mathrm{m}$ & Blob size \\
        \qquad$N_{r,f}$ & $21$ & & Points on flagellum \\
        \bottomrule
    \end{tabular}
\end{table}

The reduced-order model uses 25 cell-body points and 21 flagellum points, for a total
of 46 force points, compared with 1046 cell-body points and 371 flagellum points, for a
total of 1417 force points, in the detailed model~\cite{shindell}. This reduced discretization was selected
from preliminary calibration tests as a sparse representation that preserved the qualitative
structure of the detailed velocity field while substantially reducing the number of force
degrees of freedom. This point count is not intended to be optimal; rather, it provides a fixed sparse geometry for assessing the constrained inverse calibration, blob-size selection, PCA compression, and multi-bacterium scalability developed in this work.

The appropriate blob size in the MRS or MIRS depends on the spatial
discretization and the geometry of the problem. Previous numerical and experimental calibration of regularized-Stokeslet
models has likewise demonstrated a strong relationship between blob size, discretization scale, and accuracy~\cite{Nguyen2025}.
Therefore, the detailed-model cell body blob size
\(\epsilon_{d,c}=0.015\,\mu\mathrm{m}\) cannot be inherited directly by the coarser
reduced-order model. A preliminary free-space analysis showed that when
a blob size is too small, each reduced-order force remains sharply
localized, producing artificially large fluid speeds near the force points. Increasing
the blob size spreads the force support, reducing the near-field peak error at the
cost of a slight increase in far-field discrepancy. From this preliminary free-space
analysis, the initial reduced-order cell-body blob size was chosen as
\(\epsilon^0_{r,c}=0.02125\,\mu\mathrm{m}\), while the reduced-order flagellar blob size
was selected as \(\epsilon_f=0.0385 \,\mu\mathrm{m}\). This flagellar blob size \(\epsilon_f\) is fixed in all
reduced-order simulations, while \(\epsilon^0_{r,c}\) is used for the qualitative
comparisons in Section~\ref{sec:topology}.
The optimized value
\(\epsilon^*_{r,c}=0.18\,\mu\mathrm{m}\), determined systematically in
Section~\ref{sec:blob}, is used for the quantitative error, PCA, force-distribution,
and multi-bacterium analyses.

\subsection{Simulation Configurations}\label{sec:configs}
The origin in both free-space and near-wall simulations is placed at the junction between the flagellum and the cell body, and the velocity field is evaluated on a \(201\times31\times99\) Cartesian grid.  The grid spacing, \(d_{\mathrm{mesh}}=0.1462\,\mu\mathrm{m}\), is uniform in the $y$ and $z$ directions, but one half of that value in the $x$-direction. For all configurations, the grid extends from
\(-15d_{\mathrm{mesh}}\) to \(15d_{\mathrm{mesh}}\) in \(y\) and from
\(-68d_{\mathrm{mesh}}\) to \(30d_{\mathrm{mesh}}\) in \(z\).
The asymmetric extent in the \(z\)-direction reflects the placement of the origin at
the flagellum--cell-body junction, with the flagellum extending substantially farther
in the negative \(z\)-direction than the cell body extends in the positive
\(z\)-direction. In the free-space simulations, the grid is centered about the origin in the
\(x\)-direction. In the near-wall simulations, the no-slip boundary coincides
with one edge of the computational domain, and the origin is positioned according
to the prescribed wall distance \(h\). Thus, in our simulations, the computational domain spans
approximately \(14.62\,\mu\mathrm{m}\), \(4.39\,\mu\mathrm{m}\), and
\(14.33\,\mu\mathrm{m}\) in the \(x\)-, \(y\)-, and \(z\)-directions,
respectively.

The fluid flow is computed at \(n=16\) evenly spaced flagellar phases for both the detailed and reduced-order models of a single bacterium. Cycle-averaged quantities
are used where indicated, while phase-specific results are reported when needed to
illustrate the temporal structure of the flow or force field.

Eleven configurations are studied in total: one free-space configuration and ten near-wall
configurations.

\begin{center}
\begin{tabular}{ll}
\toprule
\textbf{Configuration} & \textbf{Wall position \(h\) (\(\mu\mathrm{m}\))} \\
\midrule
Free space & no wall \\
Near wall & \(-0.5848,\ -0.731,\ -0.8772,\ -1.0234,\ -1.1696,\) \\
          & \(-1.7544,\ -2.193,\ -2.924,\ -4.386,\ -7.31\) \\
\bottomrule
\end{tabular}
\end{center}

The ten wall positions are defined by \(h=-k\,d_\mathrm{mesh}\), with
\(k\in\{4,\allowbreak 5,\allowbreak 6,\allowbreak 7,\allowbreak 8,\allowbreak
12,\allowbreak 15,\allowbreak 20,\allowbreak 30,\allowbreak 50\}\), covering more than an order of
magnitude in wall distance. The case \(h=-2.193\,\mu\mathrm{m}\) (\(k=15\))
is used throughout the qualitative flow comparisons as a representative near-wall
configuration, for which the gap between the bacterium and the wall is approximately
four times the cell-body radius. 
A schematic comparing the evaluation grid with the detailed- and reduced-order
model discretizations is provided in Figure 1 of Supplemental Material~\cite{SupplementalMaterial}.

\section{Results}\label{sec:results}

The results are organized as follows.
Section~\ref{sec:topology} provides a qualitative validation of the reduced-order
model and identifies localized near-field velocity errors.
Section~\ref{sec:blob} then calibrates the cell-body blob size to reduce these errors
while preserving the large-scale flow structure.
Section~\ref{sec:decay} analyzes the spatial structure of the remaining error after
calibration, distinguishing near-field, far-field, and near-wall contributions.
Section~\ref{sec:pca} examines the PCA representation of the calibrated forces.
Section~\ref{sec:forces} examines how the calibrated force distribution changes with
wall distance.
Section~\ref{sec:efficiency} evaluates the computational gains of the reduced-order
model and demonstrates its use in multi-bacterium flow simulations.

\subsection{Qualitative Validation of the Reduced-Order Model}\label{sec:topology}

We first ask whether the reduced-order model preserves the qualitative structure of the high-fidelity flow field in both free space and near the wall. Here, qualitative structure refers to the large-scale streamline patterns and the boundary-induced redirection of fluid. Agreement in these features provides an important baseline validation before quantitative error analysis is performed.

\begin{figure}[!htbp]
    \centering
    \vspace{-1.25cm}
    \includegraphics[width=0.84\textwidth]{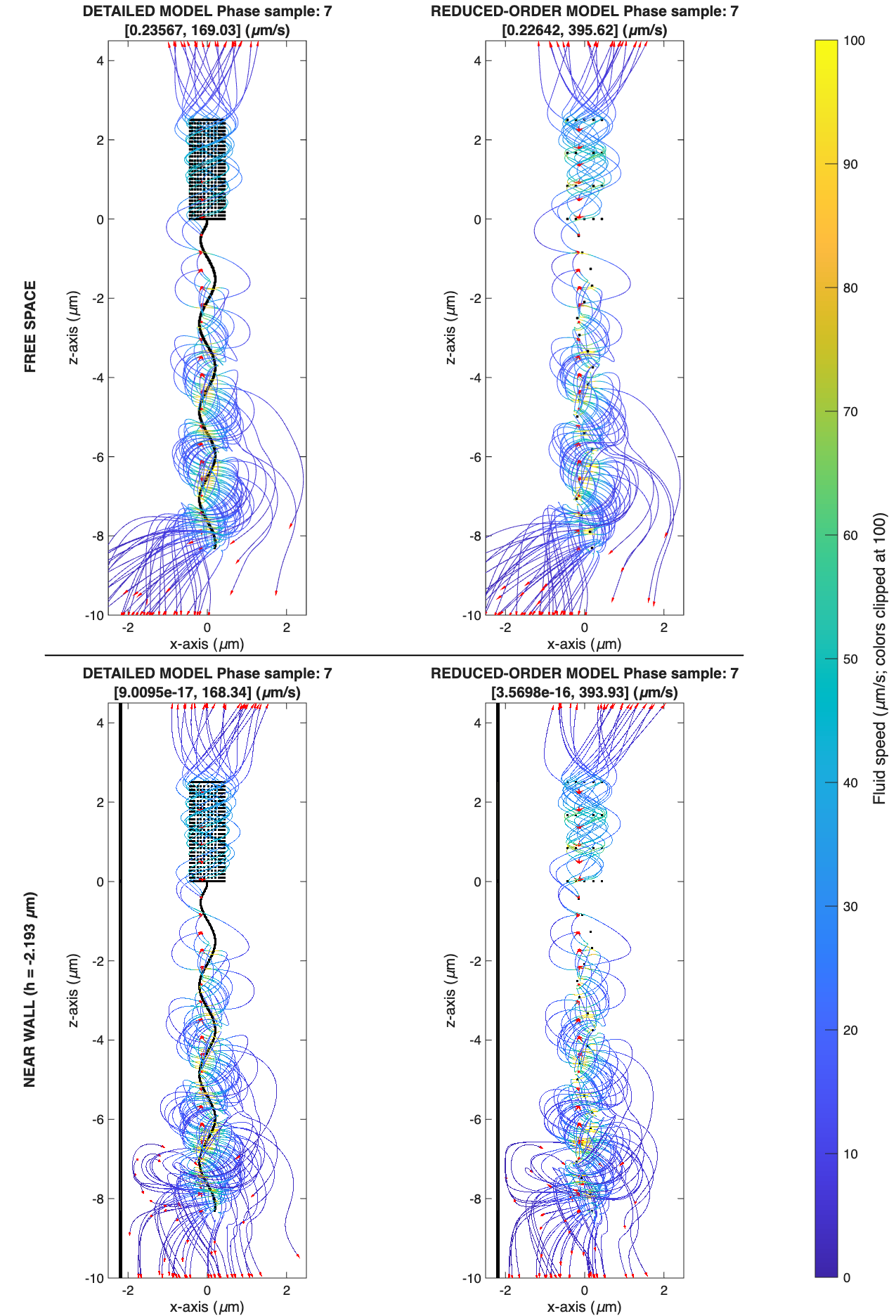}
   \caption{Instantaneous streamlines projected onto the \(xz\) plane at phase sample 7:
free space (top) and \(h=-2.193\,\mu\mathrm{m}\) (bottom), detailed (left) and reduced
order with \(\epsilon^0_{r,c}=0.02125\,\mu\mathrm{m}\) (right). Points mark the
discretizations, the solid line marks the wall, and normalized red cones indicate flow
direction. Color denotes speed on a common \(0\)--\(100\,\mu\mathrm{m}/\mathrm{s}\)
scale. The reduced model preserves the dominant structure but locally overpredicts
near-body speed.}
\label{fig:phase7}
\end{figure}

\begin{figure}[!htbp]
    \centering
    \vspace{-1.25cm}
    \includegraphics[width=0.87\textwidth]{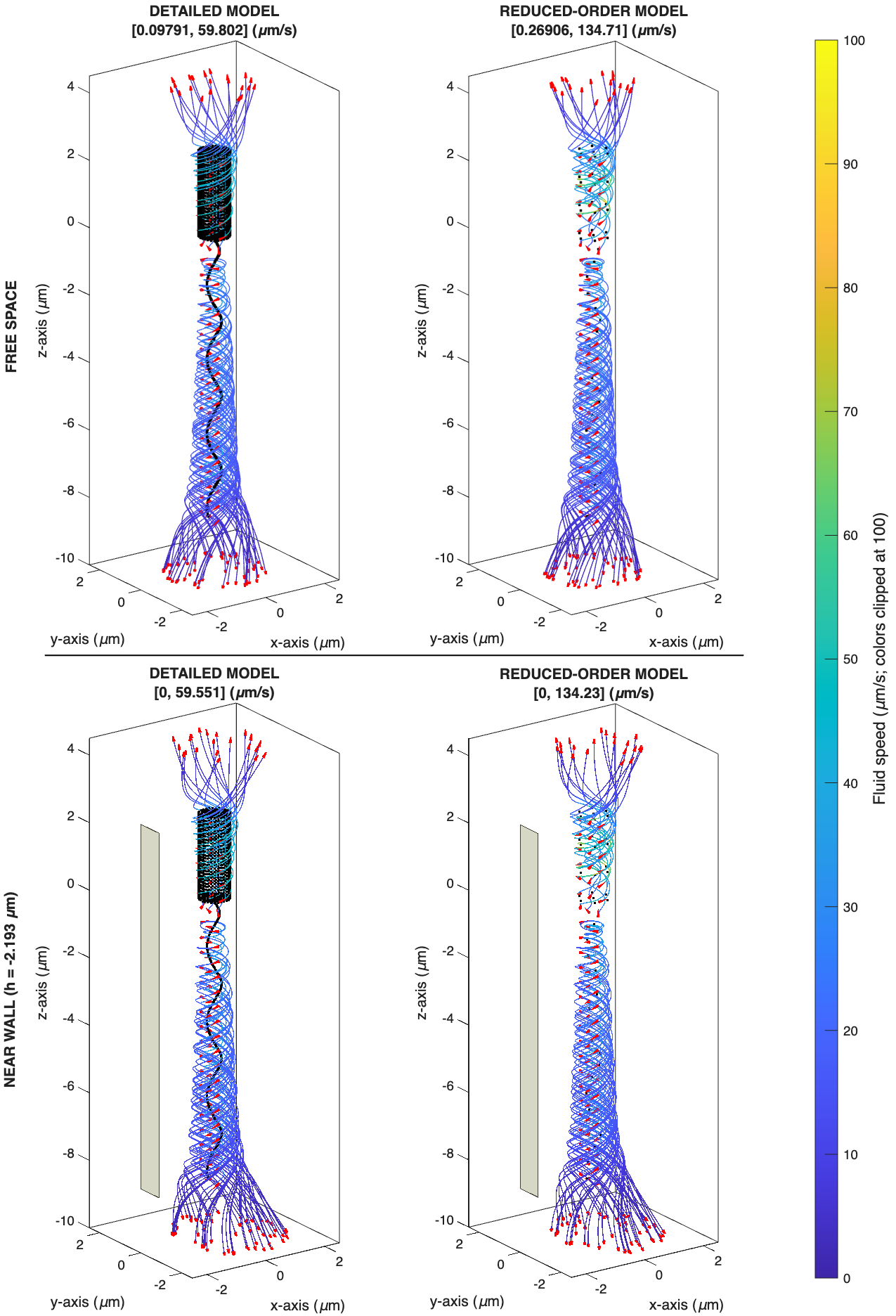}
\caption{Componentwise cycle-averaged streamlines: free space (top),
\(h=-2.193\,\mu\mathrm{m}\) (bottom), detailed (left), and reduced order
(\(\epsilon^0_{r,c}=0.02125\,\mu\mathrm{m}\); right). The swimmer geometry is calibration
phase 16, shown only for spatial reference. Bottom plane: wall; normalized red cones: flow
direction; color: speed on a shared \(0\)--\(100\,\mu\mathrm{m}/\mathrm{s}\) scale. The
dominant large-scale structure is similar.}
\label{fig:3D_avg}
\end{figure}

\begin{figure}[!htbp]
    \centering
    \vspace{-1.55cm}
    \includegraphics[width=0.68\textwidth]{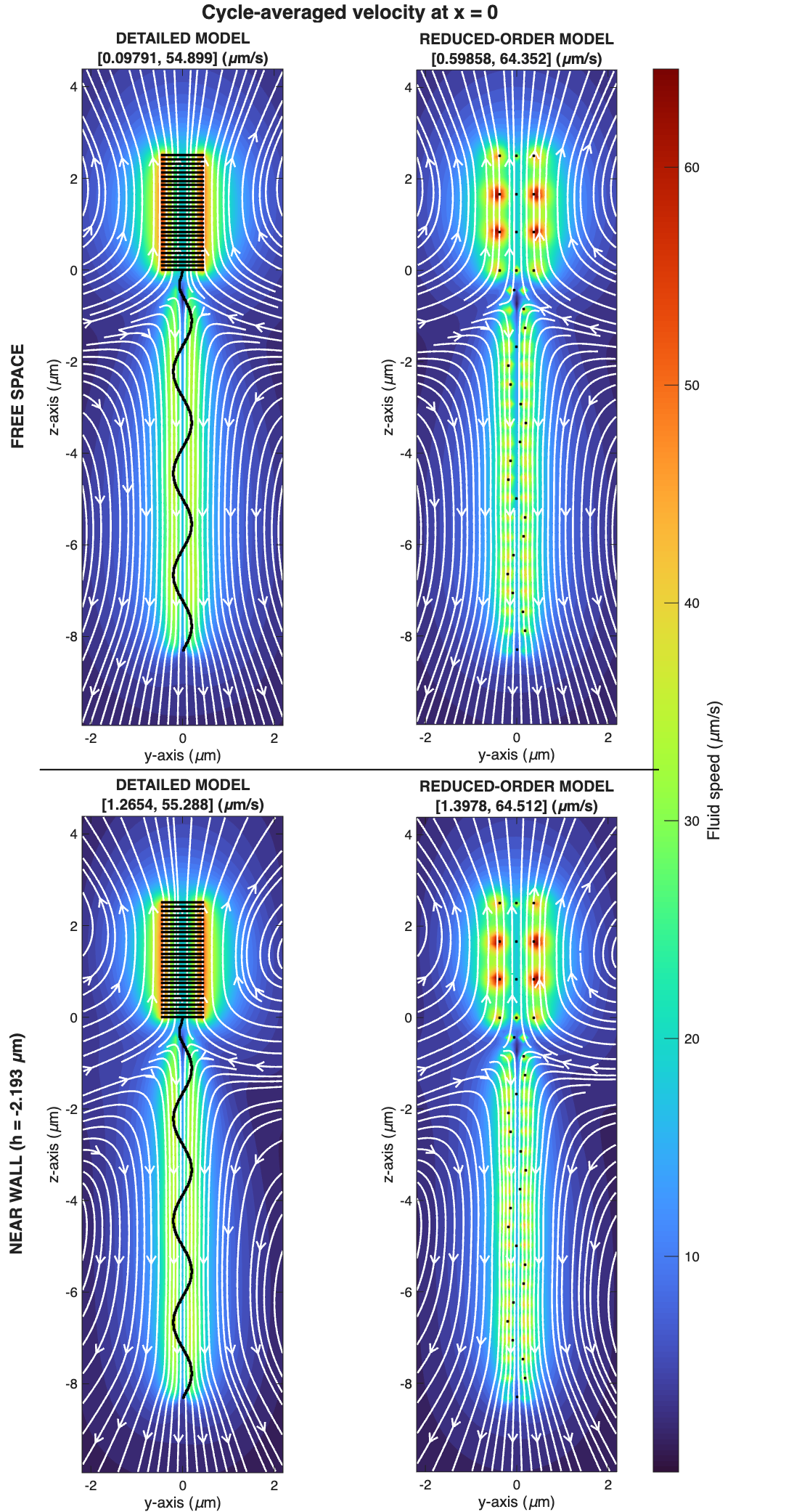}
\caption{Cycle-averaged velocity on the exact \(x=0\) plane in free space (top) and at
\(h=-2.193\,\mu\mathrm{m}\) (bottom): detailed (left) and reduced with
\(\epsilon^0_{r,c}=0.02125\,\mu\mathrm{m}\) (right). Color shows cycle-averaged speed;
streamlines show the corresponding in-plane flow.
Phase-16 geometry provides spatial reference. Dominant streamlines persist but near-body speed
hotspots motivate Section~\ref{sec:blob}.}
\label{fig:slice0}
\end{figure}

Figures~\ref{fig:phase7}--\ref{fig:slice0} compare the detailed and reduced-order
velocity fields from three complementary viewpoints using the preliminary cell-body blob
size \(\epsilon^0_{r,c}=0.02125\,\mu\mathrm{m}\). Each figure uses the same arrangement:
the top and bottom rows show free space and the near-wall configuration at
\(h=-2.193\,\mu\mathrm{m}\), respectively, while the left and right columns show the
detailed and reduced-order models. Figure~\ref{fig:phase7} presents an \(xz\) projection
of the instantaneous three-dimensional streamline field at flagellar phase sample 7,
selected to display the wall-induced flow redirection. Figure~\ref{fig:3D_avg} presents
three-dimensional streamlines of the velocity field obtained by averaging each velocity
component over all 16 phase samples. Figure~\ref{fig:slice0} shows the same
componentwise-averaged field on the exact \(x=0\) plane, which is parallel to the wall and
contains the swimmer's longitudinal axis. This figure allows localized differences between the detailed and reduced-order velocity fields near the swimmer to be compared without three-dimensional projection effects.

Across these comparisons, the reduced-order model captures the main streamline structure
of the detailed model in both free space and near the wall. In the near-wall case
\(h=-2.193\,\mu\mathrm{m}\), it captures the boundary-induced change in flow direction
within the gap between the bacterium and the wall. Within each figure, the detailed and
reduced-order panels use a shared display scale. The cross-sectional comparison in
Figure~\ref{fig:slice0} shows most directly that the dominant speed discrepancy is localized
near the cell body rather than distributed throughout the surrounding grid.

Figure~\ref{fig:slice0} shows localized speed hotspots near the projected reduced-order
cell-body points. Their location is consistent with concentration of the effective
cell-body force onto 25 reduced-order points rather than the 1046 points of the detailed
cell-body discretization. Section~\ref{sec:blob} provides further support for this
interpretation by showing that increasing the reduced-order cell-body blob size reduces
the near-field overshoot.

\begin{figure}[!htbp]
    \centering
    \includegraphics[width=0.78\textwidth]{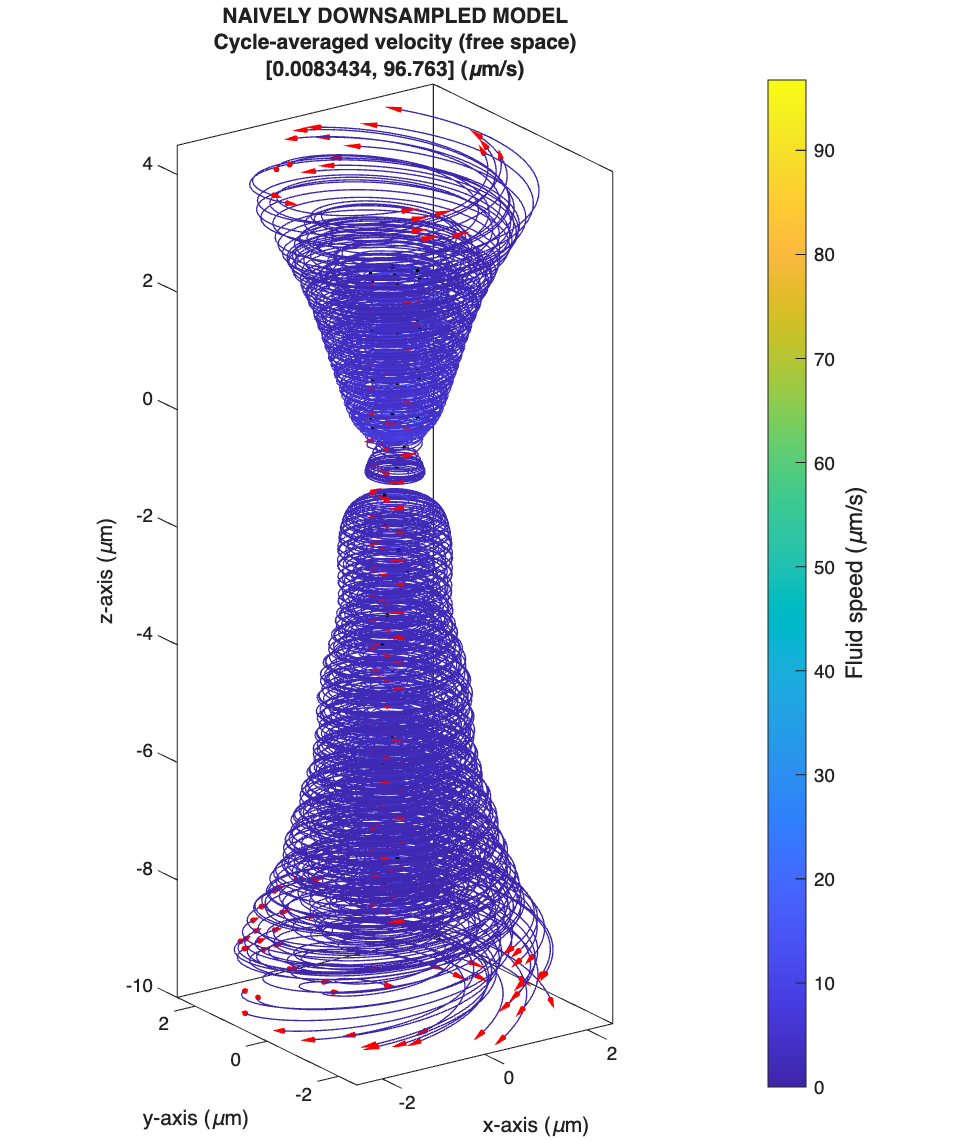}
  \caption{Cycle-averaged free-space streamlines for a naively coarsened version of the detailed model, using 25 cell-body and 21 flagellar points. At each of 16 phases, the sparse mobility problem is solved directly under the
force-free and torque-free constraints, without matching the detailed-model velocity
field. The body and flagellar blob sizes are the detailed-model values,
\(0.015\,\mu\mathrm{m}\) and \(0.025668\,\mu\mathrm{m}\), respectively. Color shows the
full computed speed range, and
normalized red cones indicate flow direction. The resulting tightly wound streamline pattern differs qualitatively
from the detailed-model result in Figure~\ref{fig:3D_avg}.}
\label{fig:bad_detailed}
\end{figure}

The agreement in Figs.~\ref{fig:phase7}--\ref{fig:slice0} raises a natural question: would simply coarsening the detailed model to the same number of force points be sufficient to preserve the detailed flow structure? To test this, we construct a naively downsampled version of the detailed model using 25 cell-body points and 21 flagellar points. The forces are obtained from the same augmented force/rigid-body mobility formulation used in the detailed model, with the detailed-model blob sizes retained, rather than by calibration against the detailed velocity field. Compared with the detailed free-space result in Figure~\ref{fig:3D_avg}, this straightforward coarse discretization produces a qualitatively different, tightly wound streamline field (Figure~\ref{fig:bad_detailed}). Thus, simply reducing the detailed-model discretization to 46 force points does not preserve its flow structure, motivating the calibrated reduced-order construction developed here.


\subsection{Calibration of the Cell-Body Blob Size}\label{sec:blob}

The near-field velocity overshoot identified in Section~\ref{sec:topology} arises from a
clear numerical mechanism: the reduced-order model represents the cell-body contribution
using only 25 calibrated force points. When the cell-body blob size \(\epsilon_{r,c}\) is too
small for this coarser discretization, the force remains too localized and produces artificially
large velocities near the reduced-order cell-body points. Increasing \(\epsilon_{r,c}\) spreads
the force support over a larger region, reducing the peak velocity without changing the
calibration procedure.

We focus this calibration on the reduced-order cell-body blob size because the
most persistent visible discrepancy is the near-field overshoot associated with
the sparse 25-point cell-body representation. The reduced-order flagellar blob
size is held fixed throughout this calibration, while only the reduced-order cell-body blob size is varied. A joint
optimization of the reduced-order cell-body and flagellar blob sizes would
require a coupled two-parameter calibration and is beyond the scope of the
present study.

\begin{figure}[!htbp]
    \centering
    \includegraphics[width=\textwidth]{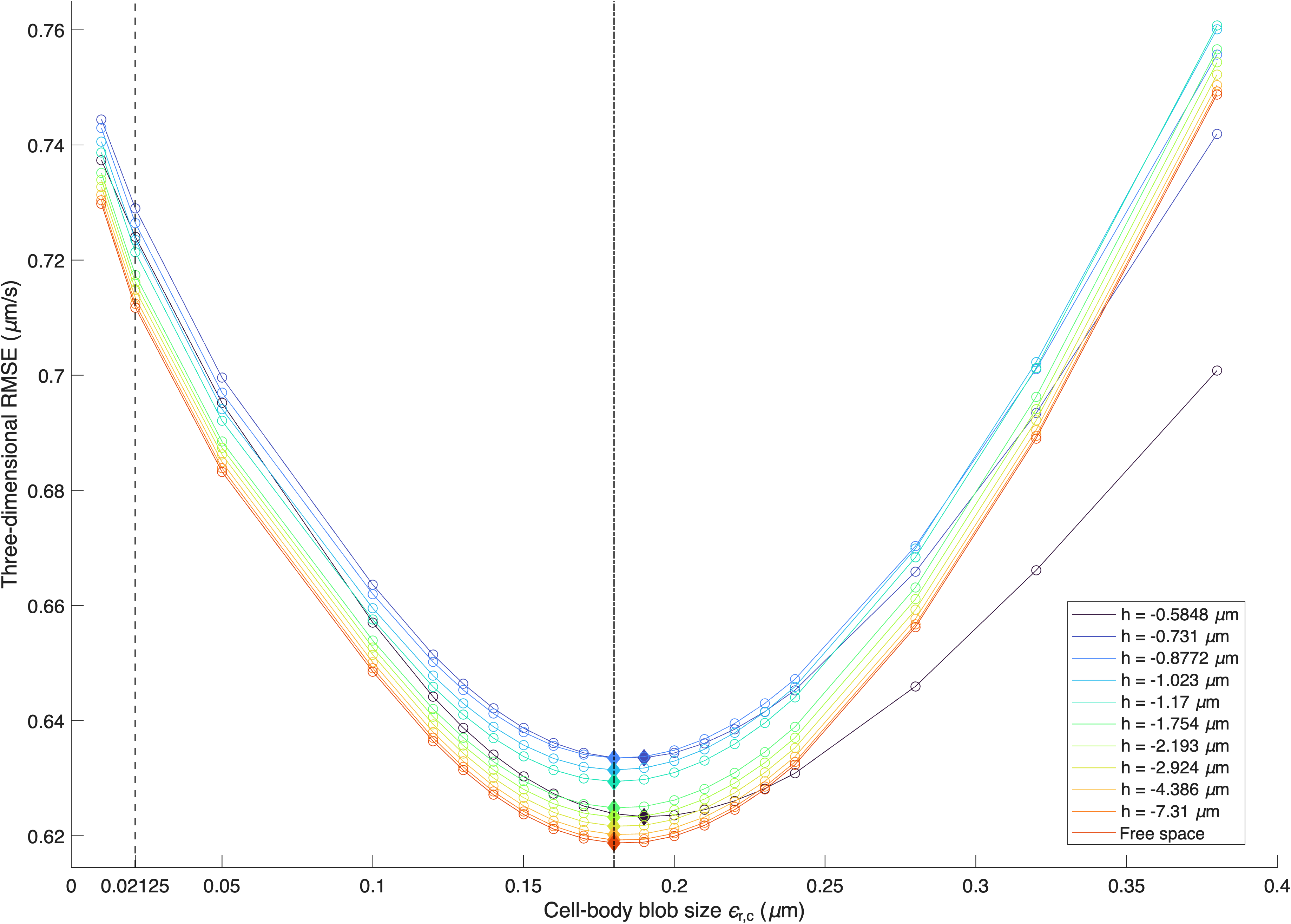}
\caption{Cycle-averaged three-dimensional speed RMSE, evaluated outside the cell body, as
a function of \(\epsilon_{r,c}\). Circles show the 20 tested blob sizes for each
configuration; diamonds identify the sampled minima: \(0.18\,\mu\mathrm{m}\) in nine
configurations and \(0.19\,\mu\mathrm{m}\) for the two closest-wall cases. The dashed and
dash-dotted lines mark the preliminary value \(0.02125\,\mu\mathrm{m}\) and the adopted
value \(0.18\,\mu\mathrm{m}\), respectively. Relative to the preliminary value, the
adopted value reduces the RMSE by \(12.73\%\)--\(13.84\%\) across all configurations. The
closest-wall case, \(h=-0.5848\,\mu\mathrm{m}\), violates the blob-size--gap condition and
is treated as a numerical stress test.}
\label{fig:blob_RMSE}
\end{figure}

The central question is whether an optimal cell-body blob size \(\epsilon^*_{r,c}\) exists and whether that
optimum depends on the distance to the wall. To answer this, we computed the three-dimensional RMSE over all exterior grid points for
20 nonuniformly spaced blob sizes \(\epsilon_{r,c}\in[0.01,0.38]\,\mu\mathrm{m}\), independently for each of the eleven configurations while holding all detailed-model parameters fixed. The blob sizes were sampled more finely between \(0.12\) and \(0.24\,\mu\mathrm{m}\), in increments of \(0.01\,\mu\mathrm{m}\), and more coarsely elsewhere in the range.
In every configuration, the RMSE curve exhibits a well-defined minimum among the sampled values,
confirming that the cell-body blob size is a meaningful calibration parameter
(Figure~\ref{fig:blob_RMSE}).

The sampled minimum occurs at \(\epsilon_{r,c}^*=0.18\,\mu\mathrm{m}\) for nine of the eleven
configurations. The only exceptions occur at the two closest wall distances,
\(h=-0.5848\,\mu\mathrm{m}\) and \(h=-0.731\,\mu\mathrm{m}\), for which the configuration-specific optimal blob size is slightly larger, \(\epsilon_{r,c}^* = 0.19\,\mu\mathrm{m}\). Even in these two cases, using the uniform value
\(\epsilon_{r,c}^*=0.18\,\mu\mathrm{m}\) increases the RMSE by only about \(0.08\%\) and
\(0.001\%\), respectively. The
sampled minima are marked in Figure~\ref{fig:blob_RMSE}; their narrow range supports the use
of a single cell-body blob size across free-space and near-wall configurations.

For this reason, we use \(\epsilon_{r,c}^*=0.18\,\mu\mathrm{m}\) in all subsequent
simulations. Relative to the preliminary value
\(\epsilon^0_{r,c}=0.02125\,\mu\mathrm{m}\), this uniform choice reduces the
three-dimensional RMSE by \(12.73\%\) to \(13.84\%\) across all eleven configurations,
with variation of at most \(\pm1\) percentage point over the full range of wall distances.
Supplemental Material, Item~2 provides corresponding qualitative flow-field and
speed-error comparisons for the preliminary and optimized blob sizes. For each
diagnostic, the result obtained with
\(\epsilon^0_{r,c}=0.02125\,\mu\mathrm{m}\) is followed immediately by the result
obtained with \(\epsilon^*_{r,c}=0.18\,\mu\mathrm{m}\), facilitating direct
comparison.

The supplementary comparisons show that the optimized blob size preserves the large-scale streamline structure while reducing speed discrepancies across the sampled cross-sections. The localization of the remaining error near the cell body, together with the weak dependence of the RMSE-minimizing blob size on wall distance, suggests that the blob-size sensitivity is primarily controlled by the concentration of force onto the 25 reduced-order cell-body points. In contrast, the MIRS image system modifies the wall-corrected flow field but appears to have only a secondary influence on the blob size that minimizes the error over the range of wall distances considered.

This observation has a practical consequence: the cell-body blob size can be calibrated once in free space
and then used across the near-wall configurations, rather than recalibrated separately at
each wall distance, provided the blob-size-gap condition
\(\epsilon_{r,c}<|h|-r\) is satisfied. This condition is violated only at the closest wall
distance, \(h=-0.5848\,\mu\mathrm{m}\), where the gap between the cell-body surface and
the wall is \(0.1448\,\mu\mathrm{m}\). Results for this closest case are therefore
interpreted with additional caution, as discussed in Section~\ref{sec:decay}. All subsequent
analyses use the optimized value \(\epsilon_{r,c}^*=0.18\,\mu\mathrm{m}\).

\subsection{Spatial Structure of the Modeling Error}\label{sec:decay}

Having identified the optimal blob size, we ask: where is the residual error concentrated
and what physical mechanisms govern its spatial decay away from the cell body?
This question matters for determining the domain of validity of the reduced-order model and for predicting
how its accuracy scales with bacterial geometry and wall distance.

The $x$-slice RMSE at multiple cross-sections perpendicular to the $x$-axis is used to
resolve the spatial structure. On every cross-section, grid points inside the cell body
are excluded from the RMSE calculation. Cross-sections intersecting the cell body
(\(|x|<r=0.44\,\mu\mathrm{m}\)) are displayed for context but excluded from the spatial-decay fits.
Because the wall (at $x=h$) and the cell body create physically distinct environments on
either side of the bacterium, the right side ($x>0.44$, facing away from the wall) and left
side ($x<-0.44$, between the bacterium and the wall) are treated separately.

\begin{figure}[!htbp]
  \centering
  \includegraphics[width=\textwidth]{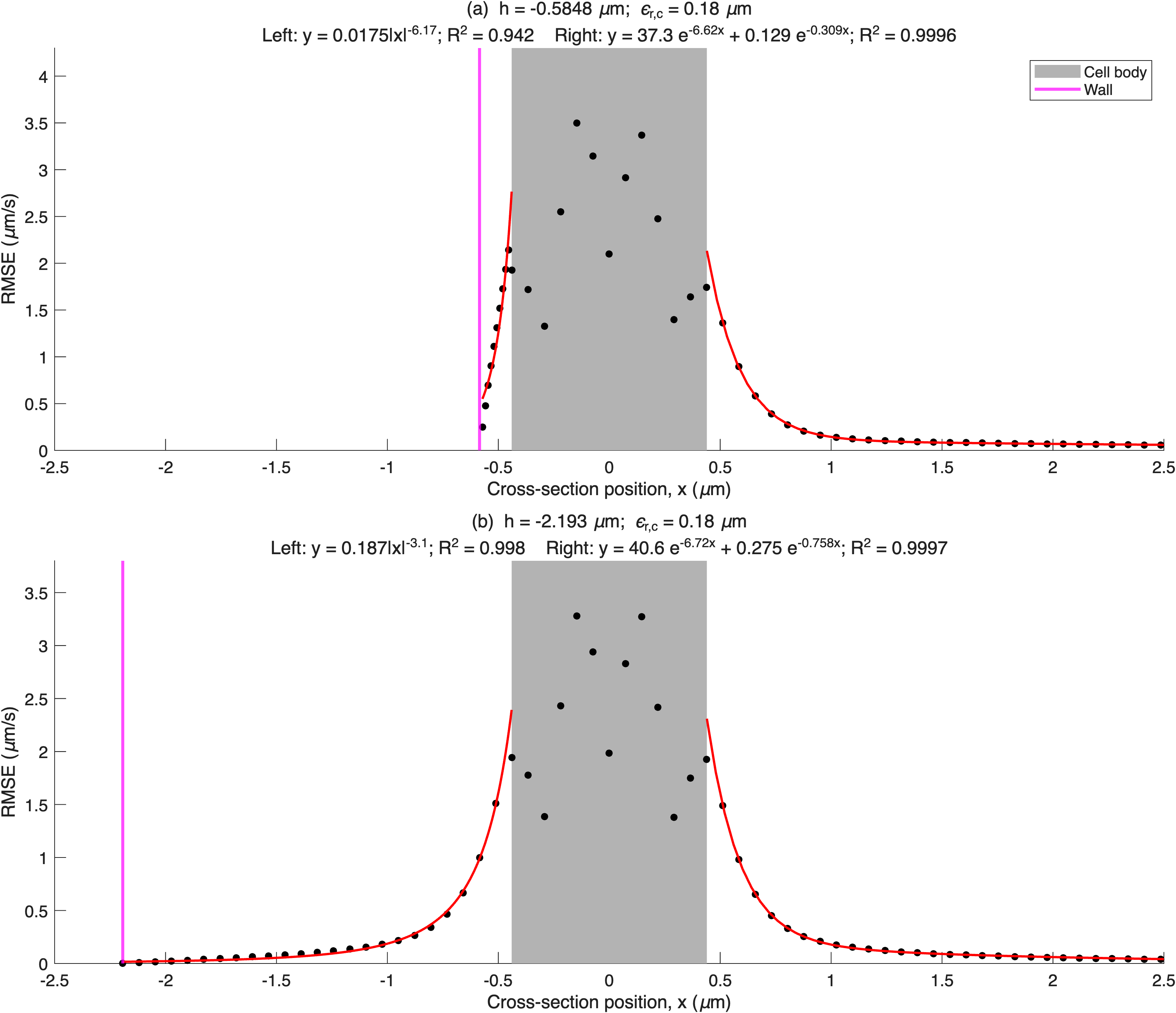}
  \caption{Cycle-averaged speed RMSE on cross-sections perpendicular to the \(x\)-axis for
\(h=-0.5848\,\mu\mathrm{m}\) (top) and \(h=-2.193\,\mu\mathrm{m}\) (bottom), using
\(\epsilon_{r,c}^*=0.18\,\mu\mathrm{m}\). The shaded interval marks cross-sections that
intersect the cell body and are excluded from the decay fits; the magenta line marks the
wall. Two-exponential and power-law curves fit the away-from-wall and wall-facing sides,
respectively. The closest-wall power-law fit is interpreted qualitatively because this
configuration violates the blob-size--gap condition.}
\label{fig:rmse_slices}
\end{figure}

Representative \(x\)-slice fits at two wall distances are presented in Figure~\ref{fig:rmse_slices}. The corresponding \(x\)-slice results for all eleven configurations, together
with the analogous \(y\)- and \(z\)-slice RMSE profiles, are provided in
Supplemental Material, Item~3.

\subsubsection{Right side: empirical two-exponential fit}

On the right side of the bacterium, where \(x>0.44\,\mu\mathrm{m}\), the
\(x\)-slice RMSE is well described by the empirical two-term exponential
\[
\operatorname{RMSE}(x)
=A_\mathrm{slow}e^{-\lambda_\mathrm{slow}x}
+A_\mathrm{fast}e^{-\lambda_\mathrm{fast}x},
\]
where \(\lambda_\mathrm{slow}\) and \(\lambda_\mathrm{fast}\) are positive decay
rates. The fit achieves \(R^2\geq0.99\) for all eleven configurations.
Figure~\ref{fig:rmse_slices} shows representative fits for the closest-wall
case, \(h=-0.5848\,\mu\mathrm{m}\), and the near-wall case
\(h=-2.193\,\mu\mathrm{m}\); fits for all eleven configurations are provided in
Supplemental Material, Item~3. A single exponential or power law gives a noticeably
poorer fit over the sampled range. The two-exponential function is therefore
used as a compact empirical representation of the rapid near-field decay and
more gradual far-field variation; its two terms are not interpreted as
uniquely identifying separate physical mechanisms.

The fast term has
\(\lambda_\mathrm{fast}\approx6.22\)--\(6.73\,\mu\mathrm{m}^{-1}\) and
\(A_\mathrm{fast}\approx33.2\)--\(40.7\,\mu\mathrm{m}/\mathrm{s}\), with only
weak wall-distance dependence. Its corresponding decay length,
\(1/\lambda_\mathrm{fast}\approx0.15\)--\(0.16\,\mu\mathrm{m}\), indicates that
this contribution is confined to the near-body region, consistent with the
localized speed hotspots around the reduced-order cell-body force points
identified in Section~\ref{sec:topology}.

The slow term represents the smaller discrepancy farther from the body and has
\(\lambda_\mathrm{slow}\approx0.28\)--\(0.77\,\mu\mathrm{m}^{-1}\) and
\(A_\mathrm{slow}\approx0.08\)--\(0.28\,\mu\mathrm{m}/\mathrm{s}\). Its decay
rate varies more with wall distance than that of the fast term, while its
amplitude remains small throughout the sampled range.

\subsubsection{Left side: empirical wall-facing power-law fit}

On the wall-facing side, the RMSE behaves differently from the right-side decay because the
evaluation region lies between the bacterium and the wall. In free space, the left and right
sides are visually symmetric, as expected in the absence of a boundary. For large wall
distances, the left-side behavior remains close to the free-space case. As the wall approaches,
however, this symmetry is progressively lost: the shrinking wall-body gap and the nearby
MIRS image system produce a steeper wall-facing error profile.

Within the gap between the wall and the cell body, the RMSE increases as the cross-section
approaches the cell-body surface (Figure~\ref{fig:rmse_slices}). Over this region, the
left-side error is represented empirically by a power law,
\[
    y = C |x|^{-\alpha}.
\]
This expression is used only as an empirical fit over the gap between the infinite wall and cell body in each configuration; it is not intended as an asymptotic error law.

For free space and the largest wall distances, the fitted exponent is close to
\(\alpha\approx 3\), consistent with the nearly symmetric free-space decay. As the wall
moves closer to the bacterium, the exponent increases monotonically, reaching
\(\alpha\approx 6.17\) at the closest wall distance \(h=-0.5848\,\mu\mathrm{m}\). Figure~\ref{fig:rmse_slices} presents
representative cases, while the complete set of left-side fits is provided in
Supplemental Material, Item~3.

Within the sampled wall--body gaps, the increasing fitted exponent provides a
quantitative measure of the increasingly sharp localization of the error near
the cell body as the wall approaches. The trend is consistent with the increasing
influence of the MIRS image system near the boundary: as the image system associated with
each regularized Stokeslet lies closer to the physical source and evaluation region, the
velocity field develops steeper spatial variation in the wall-facing gap. Because the detailed
and reduced-order models represent the force distribution at different spatial resolutions,
these stronger near-wall image contributions can amplify their local mismatch.

The closest wall cases should be interpreted with additional caution. For the five nearest
wall distances
\[
h=-0.5848,\,-0.731,\,-0.8772,\,-1.0234,\,-1.1696\,\mu\mathrm{m},
\]
the original computational grid yields only two to ten \(x\)-slices on the left side, so a
supplementary grid with twelve uniformly spaced slices from \(x=h\) to
\(x=-0.44\,\mu\mathrm{m}\) was computed to constrain the left-side fits. Even with this
refinement, the closest case remains a stress test: at
\(h=-0.5848\,\mu\mathrm{m}\), the blob-size-gap condition is violated
\[
\epsilon_{r,c}=0.18\,\mu\mathrm{m} > \mathrm{gap}=0.1448\,\mu\mathrm{m},
\]
so the fitted exponent for this case should be viewed as qualitative rather than
quantitative.

Taken together, the two-exponential fit on the right side and the power-law fit
on the wall-facing side provide a compact empirical description of the spatial
error profiles over the sampled domain. The fitted trends are consistent with
a localized discretization mismatch and increasing near-wall spatial
variation, but they are not intended to establish unique physical decay laws. The right side
separates the residual error into a rapidly decaying near-field component and a small
far-field component, while the left side shows how the wall-facing error steepens as the
bacterium approaches the boundary.

\subsection{Few-Mode PCA Representation of the Calibrated Force Field}\label{sec:pca}

For the reduced-order model to be useful in phase-dependent simulations, its
calibrated force distribution must admit a compact representation over the
flagellar cycle. If many PCA modes were required, the force representation
would lose much of its compactness and online evaluation would become less
efficient. We therefore ask how many modes are needed
to reconstruct the phase-dependent forces, and whether this mode count changes as the
bacterium approaches the wall.

As described in Section~\ref{sec:PCA}, PCA is applied separately to each
Cartesian force component \(i=1,2,3\) of the calibrated reduced-order force
ensemble. For each component, PCA provides three related outputs: eigenvalues,
which determine the relative variance captured by each mode; eigenvectors
\(\mathbf{v}_{i,k}\), which describe spatial force patterns across the 46
reduced-order model points; and phase-dependent coefficients \(B_{i,k}(p)\),
which describe how strongly each spatial pattern is expressed at flagellar
phase \(p\). Here \(k\) indexes the PCA mode for force component \(i\).

\begin{figure}[!htbp]
    \centering
    \vspace{-1.3cm}
    \hspace*{-0.1\textwidth}
    \includegraphics[width=\textwidth]{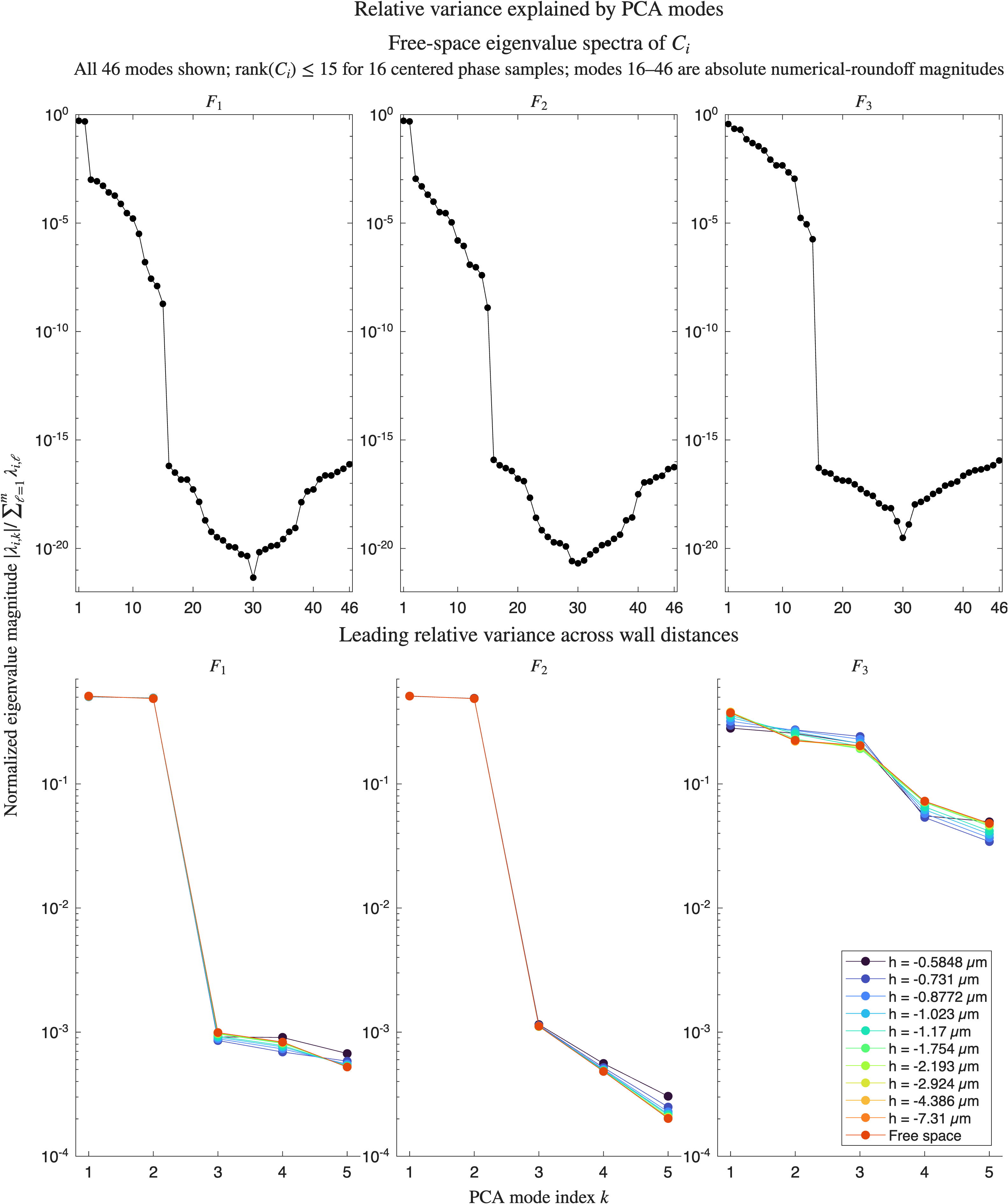}
   \caption{PCA eigenvalue spectra. Top: all 46 free-space modes. Mean-centering 16 phase
samples limits the rank to 15; modes 16--46 are zero in exact arithmetic, so their plotted
magnitudes reflect roundoff. Bottom: the first five modes across eleven configurations.
The lateral spectra drop after mode 2, whereas the axial modes are less separated. Together
with Table~\ref{tab:F3_mode_convergence}, these spectra support two modes for each lateral
component and one axial mode, denoted \(2/2/1\).}
\label{fig:variance}
\end{figure}

\begin{figure}[!htbp]
  \centering
  \vspace{-1.3cm}
  \includegraphics[width=\textwidth]{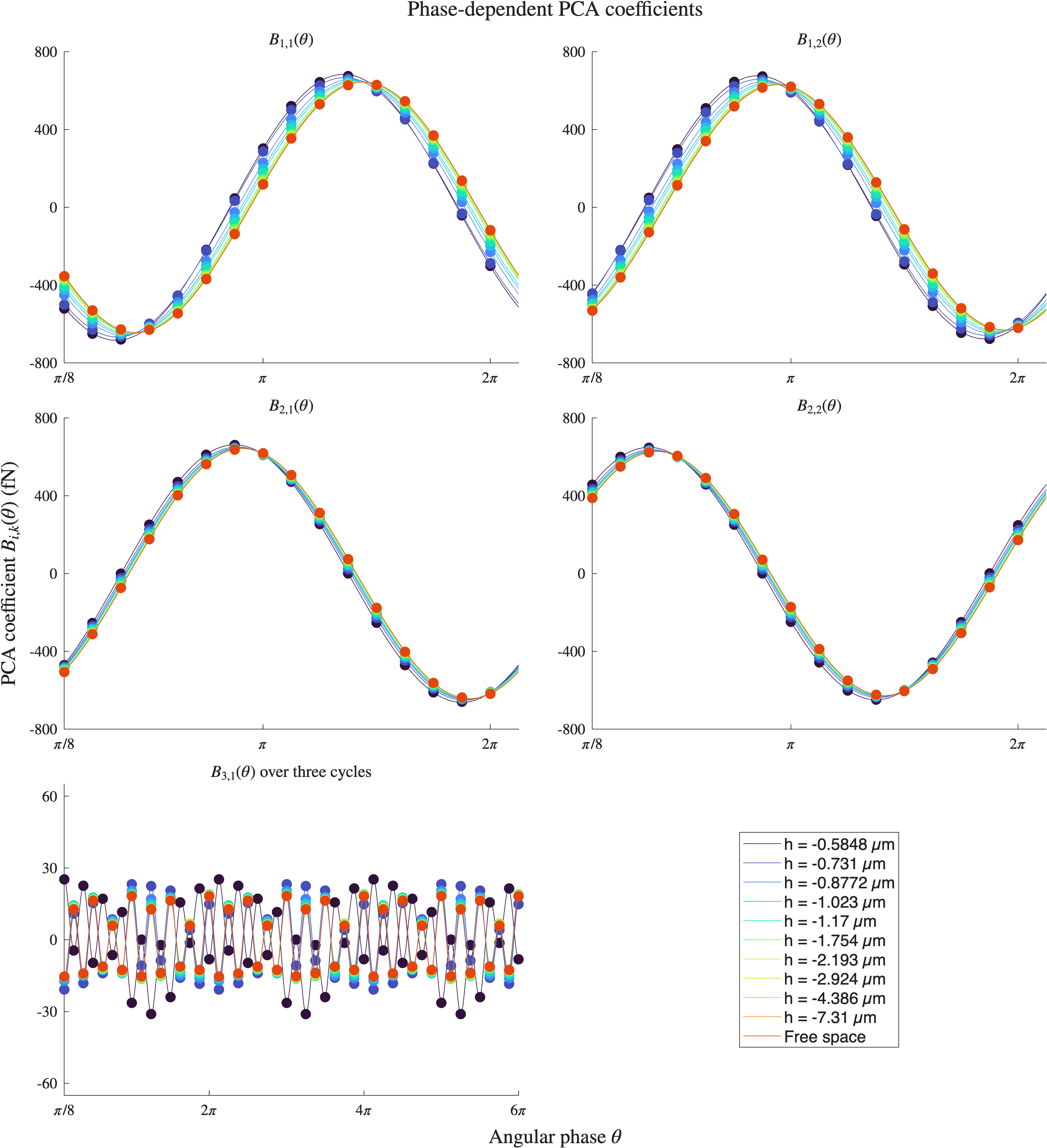}
 \caption{Phase-dependent PCA coefficients for all eleven configurations. The two retained
coefficients of \(F_1\) and \(F_2\) are represented by first-harmonic least-squares fits,
\(a\sin\theta+b\cos\theta+c\). The retained axial coefficient \(B_{3,1}\) is represented
by a periodic cubic spline whose value, slope, and curvature match across the cycle
boundary; three repeated cycles illustrate its periodic continuation. The axial
coefficient varies more rapidly with phase than the nearly sinusoidal lateral
coefficients.}
\label{fig:coefficients}
\end{figure}

\begin{figure}[!htbp]
    \centering
    \vspace{-1.3cm}
    \hspace*{-0.06\textwidth}
    \includegraphics[width=\textwidth]{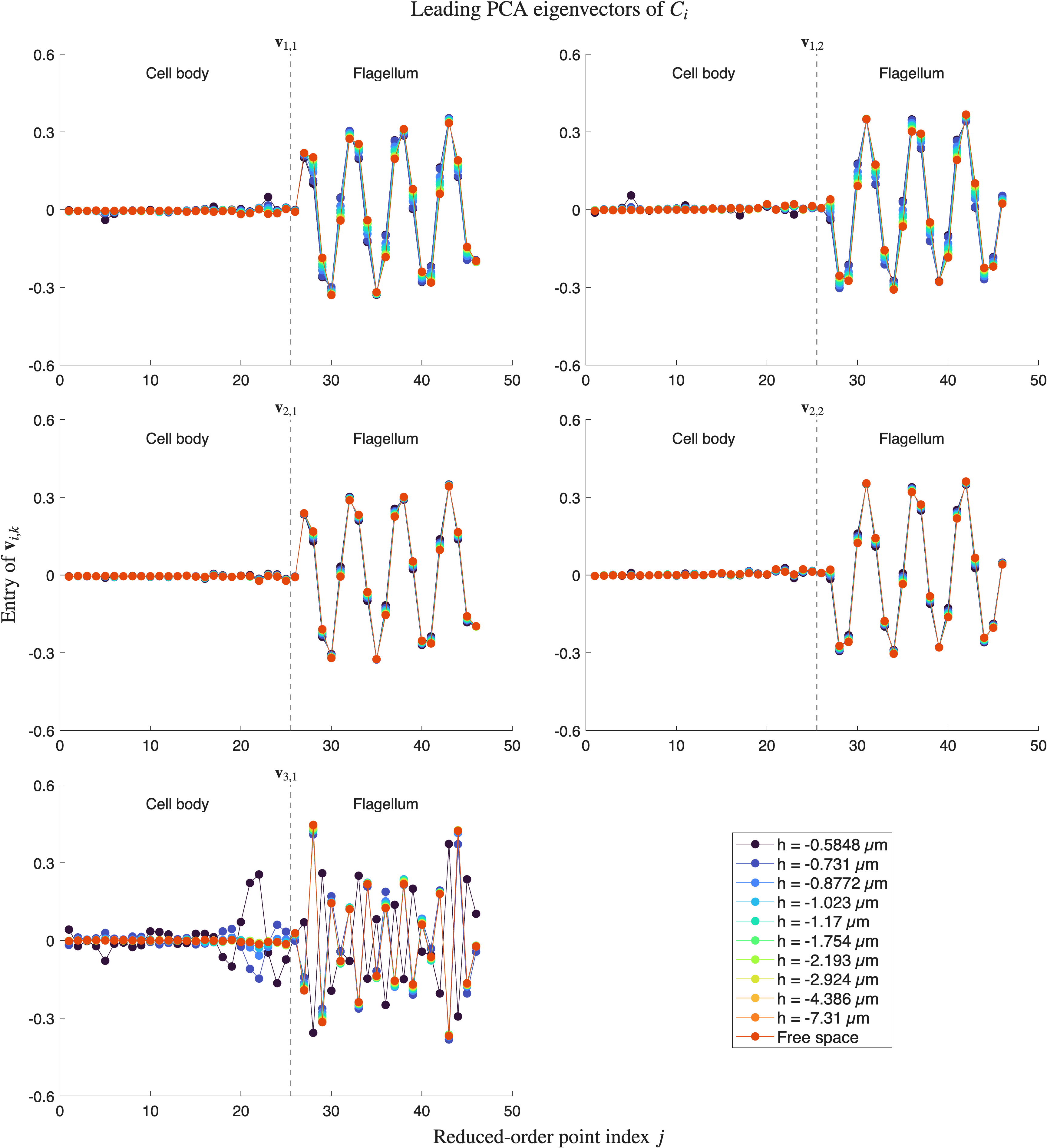}
   \caption{Leading PCA eigenvectors across all eleven configurations. The
dashed line separates cell-body indices 1--25 from flagellar indices 26--46. The lateral
modes are predominantly flagellar and vary little with wall distance. The leading axial
mode is generally flagellum-localized but mixes with the smoother, body-localized second
mode in the two closest-wall cases as their eigenvalues approach degeneracy. These cases,
particularly the closest-wall blob-size--gap stress test, are interpreted cautiously.
Supplemental Material, Item~4 compares both axial modes and reports their body/flagellum
squared-norm fractions.}
\label{fig:eigenvectors}
\end{figure}

The analysis below examines these outputs in turn. The variance spectra and
velocity-reconstruction tests determine the number of retained modes
(Figure~\ref{fig:variance} and Table~\ref{tab:F3_mode_convergence}), the
coefficient curves describe their variation over the flagellar cycle
(Figure~\ref{fig:coefficients}), and the eigenvectors show how the dominant
spatial force patterns are distributed over the cell body and flagellum
(Figure~\ref{fig:eigenvectors}). Finally, we verify that modal truncation and
continuous coefficient fitting preserve the force-free and torque-free
conditions to high numerical accuracy.

\FloatBarrier

More specifically, the eigenvalues in Figure~\ref{fig:variance} are those of the unnormalized feature covariance matrices introduced in Eq.~\eqref{eq:PCA_cov}. For each force component, an eigenvalue measures the total mean-centered squared variation across the 16 phases along a spatial force pattern defined over the 46 reduced-order points; dividing the eigenvalue by \(n-1=15\) gives the sample variance of the corresponding modal coefficient. The eigenvalues therefore quantify the phase-dependent variation of the calibrated force ensemble.

The lateral components \(F_1\) and \(F_2\), corresponding to the \(x\)- and
\(y\)-directions, are strongly dominated by two modes. For both components, the relative
variance drops by several orders of magnitude after mode 2, consistently across all eleven
configurations (Figure~\ref{fig:variance}). Thus the dominant lateral force variation is well
represented by a two-dimensional PCA subspace. The corresponding modal coefficients
\(B_{i,1}\) and \(B_{i,2}\), \(i=1,2\), vary approximately sinusoidally over the flagellar
cycle (Figure~\ref{fig:coefficients}), acting like phase-shifted sine and cosine components of
the rotating helical motion. The persistence of this sinusoidal structure across all wall
distances indicates that the dominant lateral force variation remains tied to the periodic
rotation of the flagellum, even near the boundary.

The axial component \(F_3\), corresponding to the \(z\)-direction and the swimming axis,
has a less sharply separated variance spectrum (Figure~\ref{fig:variance}).  
Nevertheless, retaining additional axial modes
produces no measurable improvement in the reconstructed velocity field.
For the three representative wall distances examined in Table~\ref{tab:F3_mode_convergence}, increasing the axial
representation from one to three modes changes the three-dimensional speed
RMSE by at most \(0.2\%\), and the change is not consistently a reduction: at
\(h=-0.5848\,\mu\mathrm{m}\) and \(h=-1.0234\,\mu\mathrm{m}\) the RMSE
increases slightly with additional axial modes, while at
\(h=-2.193\,\mu\mathrm{m}\) it is essentially unchanged.
We therefore retain one \(F_3\) mode as the most compact representation
consistent with the observed velocity accuracy. The coefficient of this
retained axial mode, \(B_{3,1}\), is represented by a periodic cubic interpolant
(Figure~\ref{fig:coefficients}), providing a continuous-phase approximation of
the axial force variation over the flagellar cycle. Unlike the approximately sinusoidal
lateral coefficients, \(B_{3,1}\) varies rapidly with phase. This temporal behavior and the
spatial localization of its associated eigenvector are distinct properties of the axial mode.

\begin{table}[!htbp]
\centering
\caption{Three-dimensional speed RMSE for reconstructions retaining one, two,
or three PCA modes for the axial force component \(F_3\). Two PCA modes are
retained for each lateral force component, \(F_1\) and \(F_2\), in all cases.
The RMSE is computed from the detailed- and reduced-order fluid-speed
magnitudes over the full grid using Eq.~\eqref{RMSE_equation}.
All results use the optimized reduced-order cell-body blob size
\(\epsilon_{r,c}^{*}=0.18\,\mu\mathrm{m}\). Retaining additional axial modes
provides no meaningful reduction in RMSE for the three representative wall
distances examined.}
\label{tab:F3_mode_convergence}
\small
\setlength{\tabcolsep}{8pt}
\begin{tabular}{lccc}
\toprule
Wall position &
\multicolumn{3}{c}{Three-dimensional speed RMSE
\((\mu\mathrm{m}/\mathrm{s})\)} \\
\cmidrule(lr){2-4}
\(h\) &
One \(F_3\) mode &
Two \(F_3\) modes &
Three \(F_3\) modes \\
\midrule
\(-4d_{\mathrm{mesh}}=-0.5848\,\mu\mathrm{m}\)
  & 0.6238 & 0.6243 & 0.6249 \\
\(-7d_{\mathrm{mesh}}=-1.0234\,\mu\mathrm{m}\)
  & 0.6314 & 0.6314 & 0.6315 \\
\(-15d_{\mathrm{mesh}}=-2.1930\,\mu\mathrm{m}\)
  & 0.6232 & 0.6232 & 0.6232 \\
\bottomrule
\end{tabular}
\end{table}

The eigenvector structure shows that the dominant lateral spatial force patterns are largely
stable across wall distances (Figure~\ref{fig:eigenvectors}), but the axial component
\(F_3\) differs qualitatively from the lateral components. For \(F_1\) and \(F_2\), the
retained eigenvectors have nearly zero cell-body components and regular oscillatory
patterns along the flagellum, with little variation across the wall distances studied.
For \(F_3\), the leading eigenvector \(\mathbf{v}_{3,1}\) is instead concentrated on the
flagellum rather than the cell body, varying rapidly in sign from one flagellar point to
the next, while the smoother, cell-body-concentrated pattern appears in the second
eigenvector \(\mathbf{v}_{3,2}\). Supplemental Material, Item~4 documents this axial-mode
structure across all eleven configurations, including the variance fractions,
body/flagellum squared-norm fractions, and phase-dependent coefficients of the two
leading modes. This ordering, in which the flagellum-dominated mode carries more variance than the cell-body-dominated mode, holds across all eleven configurations, although the two modes are most cleanly separated away from the wall. In free
space, \(\mathbf{v}_{3,1}\) is \(99.9\%\) concentrated on the flagellum and
\(\mathbf{v}_{3,2}\) is essentially pure cell body, whereas at the closest wall distance the
leading three \(F_3\) eigenvalues are nearly degenerate (\(28\%\), \(26\%\), and \(21\%\) of
the variance) and the two modes correspondingly mix cell-body and flagellum content
(\(\mathbf{v}_{3,1}\) is \(18\%\) cell body/\(82\%\) flagellum, and \(\mathbf{v}_{3,2}\) is an
almost even \(51\%/49\%\) split). These percentages are squared-norm energy
fractions, computed by summing \(v_{3,k,j}^2\) over body points 1--25 or
flagellar points 26--46 and dividing by the squared norm of the complete
eigenvector. Because the two closest wall cases show the strongest
such mixing, and because the closest case violates the blob-size-gap condition, the leading
\(F_3\) eigenvector for those configurations should be interpreted cautiously.

Because PCA is applied separately to the three Cartesian force components and
only a subset of the modes is retained, the reconstructed forces are not
algebraically guaranteed to remain exactly force-free and torque-free. We
therefore evaluated the normalized constraint residuals
\begin{equation}
R_F(\theta)
=
\frac{
\left\|\sum_{\ell=1}^{m}\widehat{\mathbf f}_{\ell}(\theta)\right\|_2
}{
\sum_{\ell=1}^{m}\left\|\widehat{\mathbf f}_{\ell}(\theta)\right\|_2
},
\qquad
R_T(\theta)
=
\frac{
\left\|\sum_{\ell=1}^{m}
\bigl(\mathbf y_{\ell}(\theta)-\mathbf x_r\bigr)
\times\widehat{\mathbf f}_{\ell}(\theta)\right\|_2
}{
\sum_{\ell=1}^{m}
\left\|
\bigl(\mathbf y_{\ell}(\theta)-\mathbf x_r\bigr)
\times\widehat{\mathbf f}_{\ell}(\theta)
\right\|_2
}.
\label{eq:pca_constraint_residuals}
\end{equation}
Supplemental Material, Item~5 provides the residual definitions, physical units, and
configuration-specific results. We evaluated the reconstructed constraints at
the 16 calibration phases, 16 out-of-sample midpoint phases, and 1,001 phases
spanning the full cycle. Across these checks, the normalized force residual
remained at numerical precision and the normalized torque residual did not
exceed \(1.25\times10^{-3}\), or \(0.13\%\). The largest dense-cycle torque
residual occurred in the closest-wall stress test; excluding this case reduced
the dense-cycle maximum only slightly, from \(1.24\times10^{-3}\) to
\(1.18\times10^{-3}\). Over the dense-cycle evaluation, the
configuration-wise maximum absolute net-force residuals ranged from
\(3.94\times10^{-13}\) to \(9.22\times10^{-13}\,\mathrm{fN}\), consistent
with numerical roundoff. The corresponding absolute net-torque residuals
ranged from \(13.4\) to \(25.8\,\mathrm{fN}\,\mu\mathrm{m}\). Although these
values are not at roundoff, they represent only \(0.068\%\)--\(0.124\%\) of the sum of the magnitudes of the torque contributions from the 46 reduced-order force points. Thus, modal
truncation and continuous coefficient fitting preserve the force-free
condition and closely approximate the torque-free condition throughout the
cycle.

Together, the variance spectra and constraint-residual checks across all eleven
configurations, along with comparisons of the three-dimensional speed RMSE
obtained using one, two, and three axial modes at three representative wall
distances, support retaining two modes for each lateral force component and one
mode for the axial force component. We refer to this choice as the \(2/2/1\)
mode selection, where the three numbers give the modes retained for \(F_1\),
\(F_2\), and \(F_3\), respectively.

\subsection{Wall-Induced Changes in the Reduced-Order Force Distribution}\label{sec:forces}

\begin{figure}[!htbp]
  \centering
  \includegraphics[width=0.88\textwidth]{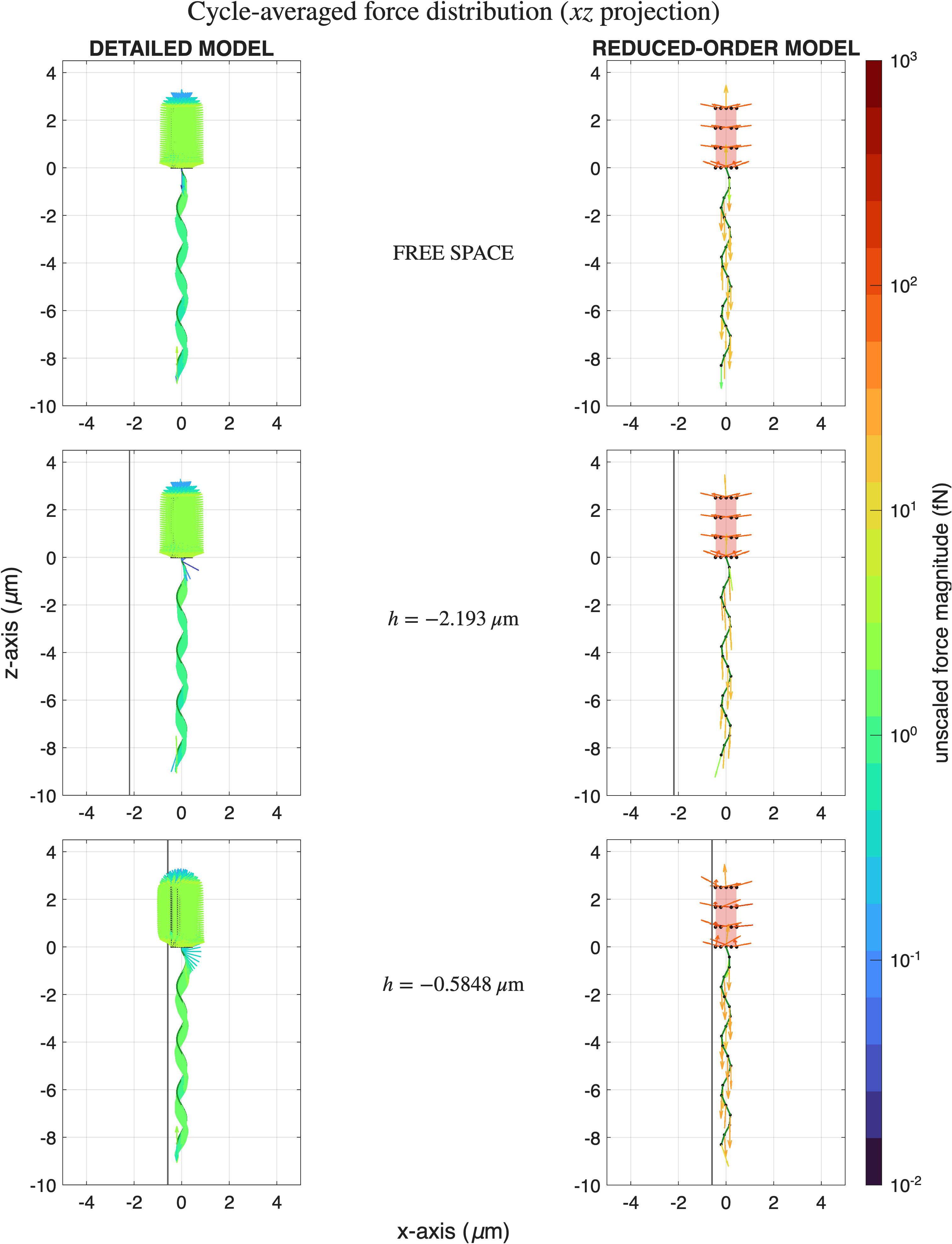}
 \caption{Cycle-averaged force vectors on the cell body and flagellum, projected onto the
\(xz\)-plane, for the high-resolution (left) and reduced-order (right) models. The solid vertical line represents the wall position in the lower four panels with \(h=-2.193\,\mu\mathrm{m}\) (middle), and
\(h=-0.5848\,\mu\mathrm{m}\) (bottom). Before projection onto the \(xz\)-plane, vectors are normalized to fixed display lengths of
\(0.78\,\mu\mathrm{m}\) (detailed) and \(0.96\,\mu\mathrm{m}\) (reduced). The colorbar gives the unscaled 3D mean force magnitude in \(\mathrm{fN}\) on a logarithmic
scale.}
\label{fig:forces}
\end{figure}

Section~\ref{sec:decay} characterized the reduced-order model through the velocity field it
produces, while Section~\ref{sec:pca} showed that the phase-dependent calibrated forces admit
a low-dimensional PCA representation. We now examine the calibrated forces directly to
understand how cycle-averaged force distributions vary as the bacterium approaches the wall. Figure~\ref{fig:forces} compares free space with the representative near-wall
configuration $h=-2.193\,\mu\mathrm{m}$ and the closest-wall configuration
$h=-0.5848\,\mu\mathrm{m}$. The solid vertical line in each near-wall panel gives the wall position at $x=h$.

For each corresponding material-point index \(\ell\), the plotted cycle-mean force is
\[
\left\langle\mathbf f_\ell\right\rangle
=\frac{1}{16}\sum_{p=1}^{16}\mathbf f_{\ell,p}.
\]
The flagellar point force  positions vary with phase, so we plot them on the flagellar geometry at \(\theta=0\) in Fig. \ref{fig:forces}. The arrow lengths are scaled differently because the
two discretizations have very different force
scales, so they are used only to show the direction of the
\(xz\)-projection. The full three-dimensional magnitude
\(\|\langle\mathbf f_\ell\rangle\|_2\) is plotted using a single logarithmic scale in
\(\mathrm{fN}\). The common color scale makes the larger reduced-order forces evident and the arrows provide the directional structure.

In free space the $(xz)$-projected cycle-averaged reduced-order body forces form a
mostly symmetric left--right pattern. The force  distribution at
\(h=-2.193\,\mu\mathrm{m}\) remains similar overall but shows differences on the body and flagellum. At
\(h=-0.5848\,\mu\mathrm{m}\), where the solid wall line lies close to the body surface, the
asymmetry is stronger: the body-force directions and flagellar-force pattern are visibly
redistributed relative to free space. The detailed-model panels show a corresponding
near-wall change, while their individual force magnitudes remain smaller because the force is
spread over many more points. This redistribution is consistent with the constrained calibration responding
to the MIRS-corrected target field as the image contribution grows.
Nevertheless, the reduced forces remain global fitting parameters:
Figure~\ref{fig:forces} demonstrates adaptation with wall position, not
pointwise recovery of physical surface traction.

\subsection{Computational Scaling and Multi-Bacterium Capability}\label{sec:efficiency}

The preceding sections show that the reduced-order model captures the dominant
high-fidelity velocity structure with controlled error, whereas the same structure is not
recovered by the direct-downsampling control simulation (Figure~\ref{fig:bad_detailed}).
We now examine the computational advantages of the calibrated representation and show how these savings translate into multi-bacterium flow simulations.

For a fixed evaluation grid, the computational work and storage required to construct
the velocity-evaluation operator scale linearly with the number of source force points.
Reducing the discretization from 1,417 detailed-model points to 46 reduced-order points
therefore decreases the dominant operator assembly and storage requirements by a factor of
\(1417/46\approx 30.8\). This scaling applies only to velocity evaluation after the force
distribution has been determined and does not include the cost of the constrained
calibration problem.

For the free-space examples considered here, translational and
rotational invariance allow the reduced-order force distribution obtained from
the calibrated single-swimmer representation to be copied, translated, and
rotated to the prescribed position and orientation of each bacterium. The
transformed force points and force vectors are then assembled into a single
multi-bacterium force representation, from which the velocity field is
evaluated on a common fluid grid. Thus, the free-space single-swimmer
representation can be used to construct multi-bacterium
configurations without repeating the calibration. This reuse does not
generally extend to transformations that change a swimmer's distance or
orientation relative to a boundary, for which the reduced-order forces may
require a configuration-specific near-wall calibration. These examples demonstrate the scalability of the calibrated force
representation for prescribed multi-bacterium flow fields. However, they do not yet
represent fully coupled trajectory simulations in which swimmer positions,
orientations, and flagellar phases evolve in response to hydrodynamic
interactions. Forces, torques, or kinematics are a common modeling
closure in reduced-order fluid-structure calculations because the true
interaction-modified force distributions are generally not known in advance.
The present examples adopt that closure to test assembly and flow evaluation, but
determining how each swimmer's forces respond to the other swimmers would
require an additional coupled response model.

\begin{figure}[!htbp]
    \centering
    \includegraphics[width=\textwidth]{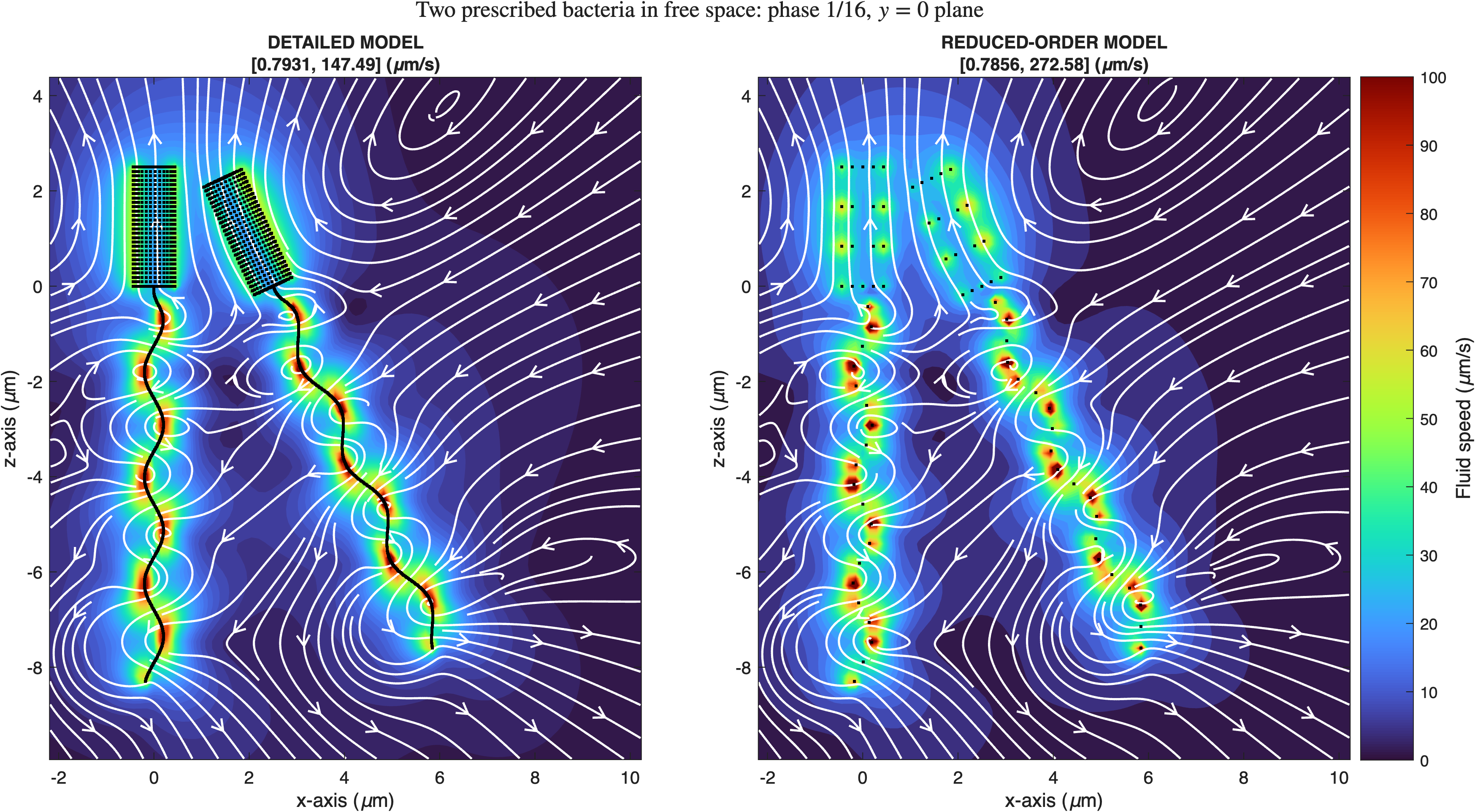}
 \caption{Instantaneous velocity on the \(y=0\) plane at phase fraction \(1/16\) for two
prescribed bacteria in free space: detailed (left) and reduced order (right). Color shows
three-dimensional speed, and streamlines show the in-plane flow. In each model, the second
swimmer's force distribution is a rigidly translated and rotated copy of the first; no
coupled two-swimmer force solve or recalibration is performed. The large-scale streamline
structures are similar, while the reduced-model near-field peaks reflect localization of
the sparse forces. Color is capped at \(100\,\mu\mathrm{m}/\mathrm{s}\); the full ranges
are reported above the panels.}
\label{fig:two_bacteria}
\end{figure}

Figure~\ref{fig:two_bacteria} compares the instantaneous velocity field on the \(y=0\)
plane at phase \(1/16\) for two bacteria in free space, computed using the detailed
and reduced-order representations. Each representation's force distribution is first
obtained for a single isolated swimmer. A second swimmer is then constructed by rotating the first swimmer's geometry and force vectors by \(-25^\circ\) about the \(y\)-axis and translating its geometry by one cell-body length in the \(x\)-direction. The combined velocity field is evaluated from the superposition of these two prescribed force distributions; no hydrodynamically coupled two-swimmer force solve or recalibration is performed. The superposed fields have broadly similar large-scale streamline structure, showing that the reduced representation can be transformed and assembled consistently despite local differences.
The main discrepancy is the localized near-field velocity differences already documented for a single bacterium (Section~\ref{sec:topology}). In instantaneous velocity fields, these localized peaks can appear near both the cell body and the flagellum, reflecting the sparse reduced-order force representation. In the cycle-averaged single-bacterium comparisons, however, the dominant
persistent overshoot is concentrated near the cell body, which is why the blob-size
calibration in Section~\ref{sec:blob} focuses on the reduced-order cell-body blob size.
In the two-bacterium case, the reduced-order representation reaches a peak speed of
\(272.6\,\mu\mathrm{m}/\mathrm{s}\), compared with \(147.5\,\mu\mathrm{m}/\mathrm{s}\) for
the detailed model. This discrepancy is therefore interpreted as a localized sparse-force
near-field artifact, rather than as a breakdown of the large-scale flow structure.

\begin{figure}[!htbp]
    \centering
    \includegraphics[width=\textwidth]{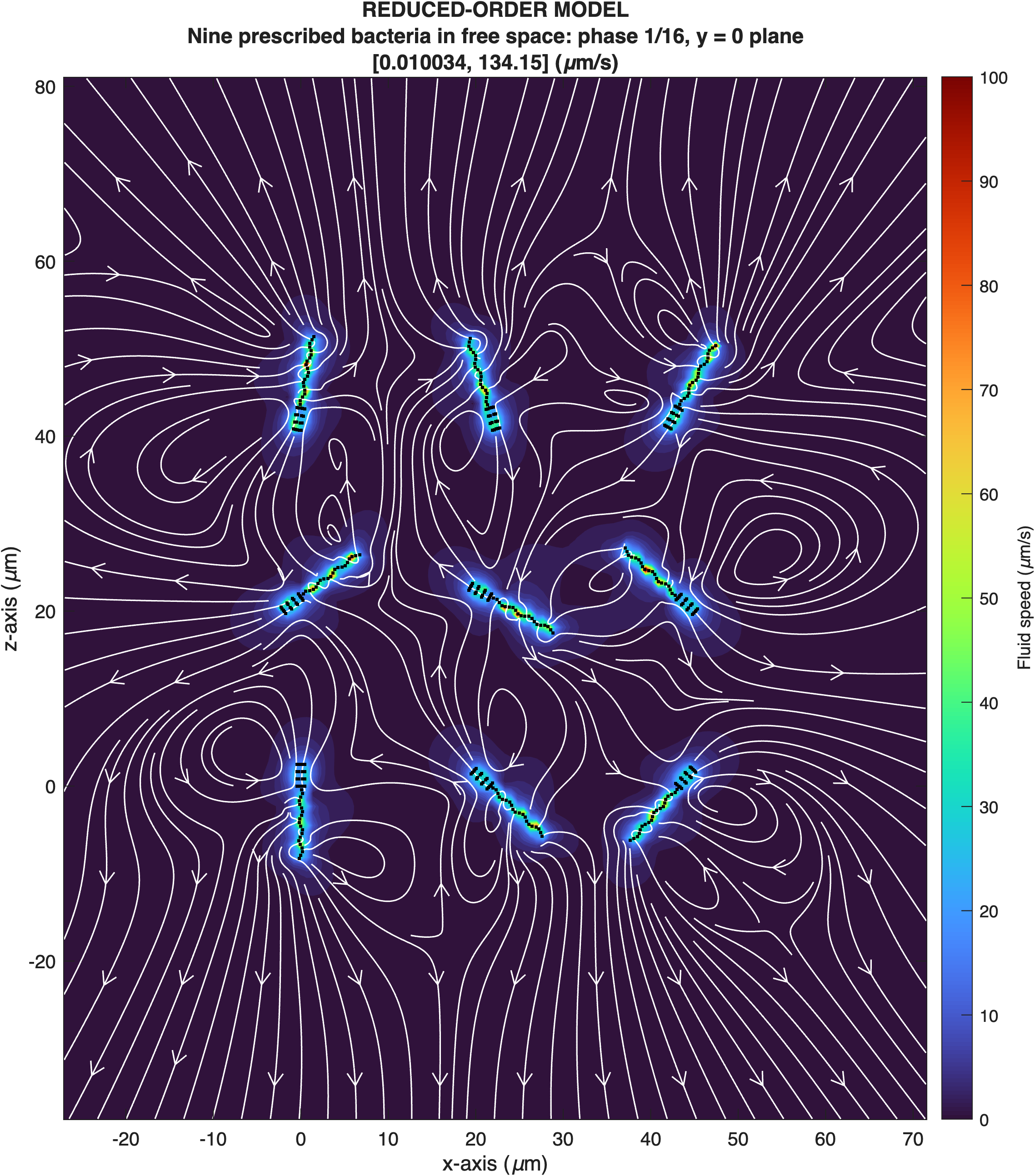}
\caption{Instantaneous velocity on the \(y=0\) plane at phase fraction \(1/16\) for nine
prescribed bacteria in free space. Color shows three-dimensional speed, and densely seeded
streamlines show the in-plane flow and resolve the near-field structure. The nine rigidly
translated and rotated single-swimmer force distributions are superposed without a coupled
multi-swimmer force solve or recalibration. Color is capped at
\(100\,\mu\mathrm{m}/\mathrm{s}\); the full range is reported above the panel.}
\label{fig:nine_bacteria}
\end{figure}

\begin{figure}[!htbp]
    \centering
    \includegraphics[width=\textwidth]{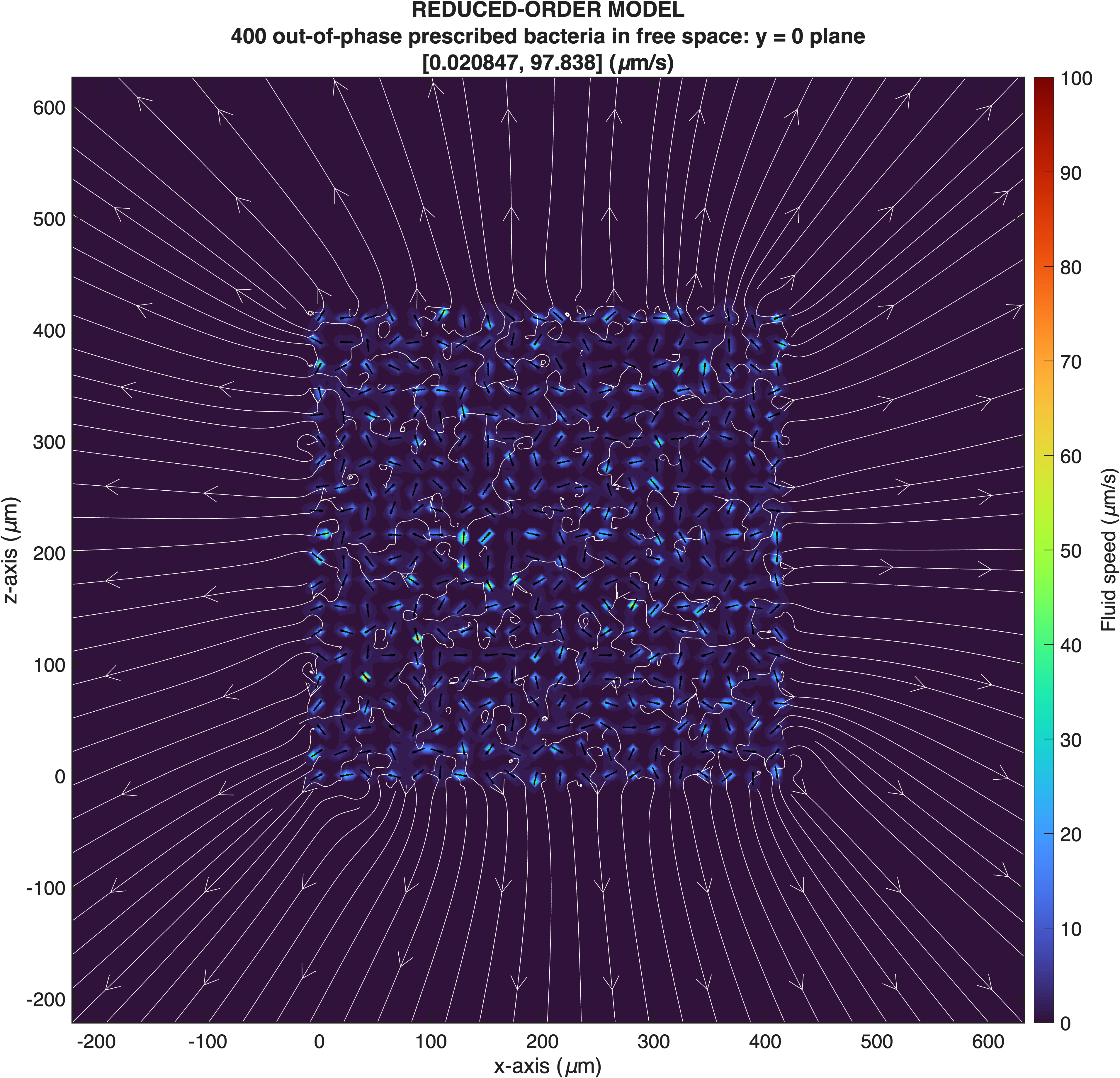}
\caption{Population-scale assembly of 400 bacteria on a \(20\times20\) lattice
in free space. Reproducible phases are drawn independently and uniformly over one cycle;
the fitted PCA coefficients reconstruct each 46-point force distribution before its
translation and rotation. On the \(y=0\) plane, color shows three-dimensional speed, and
dense streamlines show the in-plane population-scale far field. The velocity contributions
are superposed without a coupled multi-swimmer force solve, trajectories, or recalibration.
The coarse target grid resolves the far field rather than every swimmer's near field. Color
spans \(0\)--\(100\,\mu\mathrm{m}/\mathrm{s}\); the full range is reported above the panel.}
\label{fig:four_hundred_bacteria}
\end{figure}

Figures~\ref{fig:nine_bacteria} and~\ref{fig:four_hundred_bacteria} extend this 
multi-bacterium demonstration to nine and four hundred bacteria in free space, respectively.
Their purpose is to demonstrate that the reduced force representation is composable: copies
evaluated at different phases can be translated, rotated, and assembled to evaluate a common
fluid field over domains much larger than the single-swimmer calibration domain.

Figure~\ref{fig:nine_bacteria} shows the instantaneous reduced-order velocity
field on the \(y=0\) plane at phase \(1/16\) for nine prescribed bacteria in
free space. The swimmers are placed on a \(3\times3\) lattice with
\(21.6\,\mu\mathrm{m}\) spacing. Each is a translated and independently rotated
copy of the same 46-point, \(2/2/1\) single-swimmer force reconstruction. The
evaluation grid contains 22,304 target points with
\(0.731\,\mu\mathrm{m}\) spacing and spans approximately
\(98.7\times119.2\,\mu\mathrm{m}\) in the \(xz\)-plane. Dense streamline
seeding reveals the instantaneous flow around and between the swimmers produced
by the superposed prescribed force distributions.

Figure~\ref{fig:four_hundred_bacteria} demonstrates this composability at
population scale using 400 prescribed bacteria in free space. Their reference
positions form a \(20\times20\) lattice with \(21.6\,\mu\mathrm{m}\) spacing,
spanning \(410.4\,\mu\mathrm{m}\) in both lattice directions. Each swimmer is
assigned a reproducible pseudorandom phase drawn uniformly over one flagellar
cycle and a reproducible pseudorandom orientation about the \(y\)-axis. The 46-point geometry is
generated at the assigned phase, and its \(2/2/1\) force distribution is
reconstructed from the continuous PCA coefficient fits (the harmonic fits for
\(F_1\) and \(F_2\), and the periodic cubic spline for \(F_3\)) before being
rotated and translated to obtain a given configuration.

The assembled population contains 18,400 force points. On the \(y=0\) plane,
the padded evaluation domain spans approximately
\(853.8\times848.0\,\mu\mathrm{m}\) and contains 21,462 target points with
\(5.848\,\mu\mathrm{m}\) spacing. The velocity is evaluated in blocks of 242
target points, limiting the transient Stokeslet-matrix allocation to
approximately 320 MB. Because the target grid is coarse relative to an
individual swimmer, the figure displays the population-scale far field rather
than resolving every near field. It is therefore a scalability and assembly
demonstration, not an accuracy comparison with a detailed population
calculation or a hydrodynamically coupled model of collective dynamics.

\section{Conclusions}\label{sec:conclusion}

This work develops a calibrated force-representation framework for
regularized-Stokeslet simulations of force-free and torque-free microswimmers,
demonstrated using an idealized single-helix bacterial geometry in free space and near a
planar no-slip boundary. Rather than obtaining the reduced model by geometric downsampling
alone, we prescribe a sparse 46-point geometry and determine its force values
through a constrained inverse problem that approximates the velocity field of
a 1,417-point detailed model. The force-free and torque-free conditions are
enforced at each of the 16 calibration phases. The resulting representation
reproduces the principal large-scale flow structure while reducing the source
count by a factor of \(1417/46\approx30.8\).

The adopted uniform reduced-order cell-body blob size,
\(\epsilon^*_{r,c}=0.18\,\mu\mathrm{m}\), produces the lowest sampled
three-dimensional RMSE in free space and at eight wall distances. At the two
closest walls, the sampled minimum occurs at \(0.19\,\mu\mathrm{m}\), and adopting
\(0.18\,\mu\mathrm{m}\) increases the RMSE only negligibly. Relative to the preliminary
value, the adopted uniform value reduces the RMSE by
\(12.73\%\)--\(13.84\%\) across all eleven configurations. The weak variation of the
sampled minimum with wall distance suggests that blob-size sensitivity is governed primarily by the
local concentration of force on the 25 reduced-order cell-body points. The
closest-wall configuration violates the blob-size--gap criterion and is
therefore retained only as a numerical stress test. After optimization, the
largest discrepancies remain localized near the cell body, while the residual
error shows rapid near-field decay, weaker far-field variation, and a
wall-facing profile that steepens as the boundary approaches. The fitted decay
terms are empirical descriptions and are not assigned unique physical
mechanisms.

The phase-dependent calibrated forces admit a compact \(2/2/1\) PCA
representation: two modes are retained for each lateral force component and
one for the axial component. This choice is supported by the variance spectra,
mode-count sensitivity of the three-dimensional speed RMSE, and reconstructed
constraint residuals. First-harmonic fits describe the retained lateral
coefficients, while a periodic cubic spline represents the axial coefficient,
providing continuous evaluation over the flagellar cycle. The resulting force
reconstruction preserves zero net force to numerical precision and keeps the
normalized net-torque residual below \(0.13\%\) over every configuration and
phase set examined. Near the closest wall distances, the leading axial modes
mix cell-body and flagellar content, and the leading eigenvalues become nearly
degenerate in the closest case; individual eigenvectors in this regime should
therefore be interpreted cautiously.

The cycle-averaged force distributions provide complementary evidence that the
calibrated representation responds to the boundary. The comparatively
symmetric free-space pattern is progressively redistributed over the cell body
and flagellum as the wall approaches, with the strongest asymmetry in the
closest-wall stress test. These reduced-order forces are global fitting
parameters, not pointwise surface tractions; their redistribution indicates
adaptation to the wall-corrected target velocity field rather than pointwise
recovery of the detailed force distribution.

The computational benefit follows directly from the smaller source set: for a fixed collection of target points, the dominant velocity-evaluation work and operator-storage requirements decrease in proportion to the source-count reduction. A control
model obtained by direct downsampling to the same number of points, while retaining the
detailed-model blob sizes, produces a qualitatively different flow field. Thus, reducing
the point count alone does not reproduce the detailed flow structure. The two-, nine-, and
four-hundred-bacterium examples further show that independently prescribed
single-swimmer force distributions can be translated, rotated, and assembled
over increasingly large domains. These examples demonstrate composability and
scalability; they neither validate a detailed many-swimmer solution nor model
how the force distribution, rigid-body motion, or trajectory of one swimmer
responds to the others. Linearity also permits forces calibrated at one motor
frequency to be rescaled for another frequency when the phase-dependent
geometry remains unchanged.

The present near-wall construction uses a separate calibrated representation
at each sampled wall distance. A natural next step is a two-parameter model
\(\mathbf F_{\mathrm{PCA}}(p,h)\) that varies continuously with both phase and
wall position. Additional calibration samples will likely be required in the
closest-wall regime, where the force representation changes most rapidly.
Such a model should be validated at phases and wall distances excluded from
calibration, including its velocity field, constraint residuals, and rigid-body
velocities. Extending prescribed population assembly to predictive collective
dynamics will additionally require a coupled response model and systematic
many-body validation.

Although demonstrated for bacterial swimming, the calibration strategy is not
organism-specific. It can be applied to other microswimmers for which a
high-fidelity reference velocity field and the relevant physical constraints
are available. More broadly, the results show that constrained inverse
calibration, followed by phase-aware compression, offers a practical
alternative to geometric downsampling for constructing compact
regularized-Stokeslet force models.

\begin{acknowledgments}
This work was supported by the NSF Physics of Living Systems program through
awards PHY-2210609 (O.S., F.H., and H.N.) and PHY-2210610 (B.R.), and by the
Joint DMS/NIGMS Initiative to Support Research at the Interface of the
Biological and Mathematical Sciences through awards DMS-2054333 (R.C.) and
DMS-2054259 (H.N.). W.W. was supported by award DMS-2054259. Computational
resources for this work included Trinity University’s High Performance Scientific
Computing Cluster, supported by NSF MRI award ACI-1531594.
\end{acknowledgments}

\section*{Author Contributions}
H.N., W.W., and R.C. conceptualized the study. H.N. developed the free-space
model, and W.W. extended it to configurations near a planar wall, performed
the simulations and data analysis, and developed the MATLAB implementation
under the supervision of H.N. H.N. and W.W. prepared the original manuscript.
O.S. and B.R. provided feedback that clarified ambiguous terminology and technical
descriptions in the manuscript drafts. R.C. provided references to current
modeling literature and reviewed technical details. F.H. provided the biological motivation
for the bacterial-population simulations. All authors contributed to reviewing
and editing the manuscript and approved the final version.

\section*{Competing Interests}

The authors declare no competing interests.

\section*{Data and Code Availability}


The MATLAB implementation supporting this study is currently maintained in a private GitHub repository and will be made publicly available upon publication. During peer review, the code is available to the journal editors and reviewers upon request.

\clearpage
\appendix
\section{Out-of-Sample Phase Validation}
\label{app:out_of_sample_validation}

We assess the continuous phase representation separately in free space and at
the representative near-wall distance \(h=-2.193\,\mu\mathrm{m}\). Both
configurations use the same calibration and test phases and the same error
definitions, but each uses its own detailed-model solutions and calibrated
reduced-order force representation.

Let \(p\in[0,1)\) denote phase fraction and \(\theta=2\pi p\) the corresponding
angular phase. The 16 calibration samples are
\begin{equation}
 p_k^{\mathrm{cal}}=
 \begin{cases}
 k/16, & k=1,\ldots,15,\\
 0, & k=16,
 \end{cases}
 \qquad
 \theta_k^{\mathrm{cal}}=2\pi p_k^{\mathrm{cal}},
\qquad k=1,\ldots,16.
\end{equation}
The final stored calibration sample corresponds to one completed cycle and is therefore
represented as \(p_{16}^{\mathrm{cal}}=0\), the cycle boundary. The 16 out-of-sample test
phases are the interleaved midpoints
\begin{equation}
\label{eq:out_of_sample_phases}
 p_k^{\mathrm{test}}=\frac{k-\tfrac{1}{2}}{16},
 \qquad
 \theta_k^{\mathrm{test}}=2\pi p_k^{\mathrm{test}},
\qquad k=1,\ldots,16.
\end{equation}
Each test phase lies halfway between two consecutive calibration phases, with
the first midpoint lying between the cycle boundary and the sample at
\(p=1/16\). None of the test phases is included in the calibration set.

An independent detailed-model mobility problem was solved at every test phase;
none of the resulting midpoint velocity fields was used for calibration, PCA,
or coefficient fitting. For the calibration-phase baseline, the reduced forces
were reconstructed using the projected discrete PCA coefficients at the 16
sampled phases. At the test phases, the retained lateral coefficients were
evaluated from
\(B_{i,k}(\theta)=a_{i,k}\sin\theta+b_{i,k}\cos\theta+c_{i,k}\), where
\(i=1,2\), \(k=1,2\), and the three parameters were determined by least
squares from the 16 calibration values of each coefficient. The retained axial
coefficient was evaluated from the periodic cubic spline that matches value,
slope, and curvature at the cycle boundary. The discrete reconstruction at the
calibration phases provides a baseline for the error associated with the sparse
spatial representation and \(2/2/1\) PCA truncation. The midpoint calculation
additionally tests the continuous coefficient fits at 16 interleaved phases
that were excluded from calibration, PCA construction, and coefficient fitting.

The cell-body blob sizes used in the detailed and reduced-order models were
\(0.015\,\mu\mathrm{m}\) and \(0.18\,\mu\mathrm{m}\), respectively. The
corresponding flagellar blob sizes used in the Stokeslet matrices were
\(0.025668\,\mu\mathrm{m}\) and \(0.038502\,\mu\mathrm{m}\), respectively.
These are the same regularization parameters used in the preceding analysis.

At each phase \(k\), the detailed- and reduced-order velocity fields were
compared over the full three-dimensional grid, excluding grid
points inside the cell body, as in the preceding RMSE analysis. Let
\(j=1,\ldots,N\) index the retained grid points, let \(k=1,\ldots,16\)
index either phase set, and let \(M\in\{D,R\}\) denote the detailed or
reduced-order model. Three complementary quantities are reported.

First, the velocity components are averaged over the relevant phase set,
\begin{equation}
\label{eq:out_of_sample_average_velocity}
\overline{\mathbf{u}}_{M,j}
=
\frac{1}{16}
\sum_{k=1}^{16}
\mathbf{u}_{M,j}^{(k)},
\qquad
M\in\{D,R\},
\quad j=1,\ldots,N,
\end{equation}
and the cycle-averaged speed RMSE is
\begin{equation}
\label{eq:out_of_sample_average_rmse}
E_{\mathrm{avg}}
=
\sqrt{\frac{1}{N}\sum_{j=1}^{N}
\left(
\left\|\overline{\mathbf{u}}_{D,j}\right\|_2
-
\left\|\overline{\mathbf{u}}_{R,j}\right\|_2
\right)^2}.
\end{equation}
This is Eq.~\eqref{RMSE_equation} applied to the magnitudes of the
cycle-averaged velocity fields.

Second, the phasewise speed RMSE applies Eq.~\eqref{RMSE_equation}
independently at each phase:
\begin{equation}
\label{eq:out_of_sample_speed_rmse}
E_{\mathrm{speed}}^{(k)}
=
\sqrt{
\frac{1}{N}
\sum_{j=1}^{N}
\left(
\left\|\mathbf{u}_{D,j}^{(k)}\right\|_2
-
\left\|\mathbf{u}_{R,j}^{(k)}\right\|_2
\right)^2
}.
\end{equation}
The 16 values are summarized by their mean, minimum, maximum, and standard
deviation.

Third, because speed alone does not capture directional discrepancies, the
phasewise relative vector error is
\begin{equation}
\label{eq:relative_vector_error}
E_{\mathrm{rel,vec}}^{(k)}
=
\sqrt{
\frac{
\displaystyle\sum_{j=1}^{N}
\left\|
\mathbf{u}_{D,j}^{(k)}
-
\mathbf{u}_{R,j}^{(k)}
\right\|_2^2
}{
\displaystyle\sum_{j=1}^{N}
\left\|
\mathbf{u}_{D,j}^{(k)}
\right\|_2^2
}
}.
\end{equation}
This dimensionless field-level error is sensitive to differences in both
velocity magnitude and direction.

\subsection*{Free-space validation}

For free space, the cycle-averaged speed RMSE is
$0.61874\,\mu\mathrm{m}/\mathrm{s}$ at the calibration phases and
$0.61588\,\mu\mathrm{m}/\mathrm{s}$ at the midpoint phases. The mean
phasewise speed RMSE is $1.11370\,\mu\mathrm{m}/\mathrm{s}$ and
$1.10076\,\mu\mathrm{m}/\mathrm{s}$, respectively, and the corresponding
mean relative vector errors are $0.19431$ and $0.19244$. Thus, none of the
three aggregate errors increases at the unseen midpoint phases.

\subsection*{Near-wall validation at \(h=-2.193\,\mu\mathrm{m}\)}

At $h=-2.193\,\mu\mathrm{m}$, the cycle-averaged speed RMSE is
$0.62321\,\mu\mathrm{m}/\mathrm{s}$ at the calibration phases and
$0.62036\,\mu\mathrm{m}/\mathrm{s}$ at the midpoint phases. The mean
phasewise speed RMSE is $1.11766\,\mu\mathrm{m}/\mathrm{s}$ and
$1.10478\,\mu\mathrm{m}/\mathrm{s}$, respectively, and the corresponding
mean relative vector errors are $0.20612$ and $0.20415$. As in free space,
the midpoint errors remain close to, and slightly below, the corresponding
calibration-phase summaries.

\begin{table}[!htbp]
\centering
\caption{Out-of-sample phase validation in free space and at
$h=-2.193\,\mu\mathrm{m}$. Each configuration is evaluated at the 16
calibration phases and 16 out-of-sample midpoint phases. The cycle-averaged
speed RMSE is calculated after averaging the velocity components over each
phase set. The phasewise quantities are evaluated separately at each phase and
then summarized over the corresponding 16 phases.}
\label{tab:phase_validation_comparison}

\small
\setlength{\tabcolsep}{3pt}
\begin{tabular*}{\linewidth}{@{\extracolsep{\fill}}l l r r r r}
\toprule
Metric & Phase set & Mean/value & Min. & Max. & Std. dev. \\
\midrule
\multicolumn{6}{@{}l}{\textit{Free space}} \\
\shortstack[l]{Cycle-averaged speed RMSE\\
(\(\mu\mathrm{m}/\mathrm{s}\))}
& Calibration
& 0.61874 & -- & -- & -- \\
& Out-of-sample
& 0.61588 & -- & -- & -- \\
\addlinespace

\shortstack[l]{Phasewise speed RMSE\\
(\(\mu\mathrm{m}/\mathrm{s}\))}
& Calibration
& 1.11370 & 1.09547 & 1.14208 & 0.01489 \\
& Out-of-sample
& 1.10076 & 1.08609 & 1.11540 & 0.00978 \\
\addlinespace

Relative vector error
& Calibration
& 0.19431 & 0.19218 & 0.19832 & 0.00206 \\
& Out-of-sample
& 0.19244 & 0.19064 & 0.19463 & 0.00142 \\
\addlinespace
\multicolumn{6}{@{}l}{\textit{Near wall, $h=-2.193\,\mu\mathrm{m}$}} \\
\shortstack[l]{Cycle-averaged speed RMSE\\
(\(\mu\mathrm{m}/\mathrm{s}\))}
& Calibration
& 0.62321 & -- & -- & -- \\
& Out-of-sample
& 0.62036 & -- & -- & -- \\
\addlinespace

\shortstack[l]{Phasewise speed RMSE\\
(\(\mu\mathrm{m}/\mathrm{s}\))}
& Calibration
& 1.11766 & 1.09838 & 1.14812 & 0.01504 \\
& Out-of-sample
& 1.10478 & 1.08899 & 1.12083 & 0.00985 \\
\addlinespace

Relative vector error
& Calibration
& 0.20612 & 0.20294 & 0.21047 & 0.00226 \\
& Out-of-sample
& 0.20415 & 0.20119 & 0.20610 & 0.00147 \\
\bottomrule
\end{tabular*}
\end{table}

The phasewise errors vary only weakly within both phase sets, and no individual
midpoint exhibits a pronounced loss of accuracy. Because the midpoint and
calibration sets contain different physical phases, their small differences
should not be interpreted as an improvement caused by interpolation. Rather,
the absence of degradation indicates that the continuous coefficient fits add
no detectable error beyond the underlying reduced-order discrepancy at the
tested midpoints. Across the individual phases in both configurations and both
phase sets, the relative vector error ranges from \(0.19064\) to \(0.21047\),
corresponding to approximately \(19\%\)--\(21\%\) in the relative field norm.
The results therefore support continuous evaluation between sampled phases in
free space and at \(h=-2.193\,\mu\mathrm{m}\), but they do not establish
out-of-sample accuracy at every wall distance or for other swimmer geometries.

\clearpage
\bibliography{references}

\end{document}